\documentclass[11pt]{article}       
\usepackage{graphicx}
\usepackage{mathptmx}   
\usepackage{hyperref}  
\usepackage{subfigure}
\usepackage{lineno,graphicx,color}
\usepackage{amsmath}
\begin{document}

\title{Complete dimensional collapse in the continuum limit of a delayed SEIQR network model with separable distributed infectivity}

\author{C. P. Vyasarayani$^a$\thanks{Corresponding author. Email: vcprakash@iith.ac.in}
	\and
	Anindya Chatterjee$^b$\thanks{Email: anindya100@gmail.com, anindya@iitk.ac.in}  }

\date{$^a$Mechanical and Aerospace Engineering\\ Indian Institute of Technology Hyderabad \\
	Sangareddy, 502285, India\\ \smallskip $^b$Mechanical Engineering\\ Indian Institute of Technology Kanpur\\ Kanpur, 208016, India\\ \smallskip \today}

\maketitle

\begin{abstract}
We take up a recently proposed compartmental SEIQR model with delays, ignore loss of immunity in the context of a fast pandemic, extend the model to a network structured
on infectivity, and consider the continuum limit of the same with a simple separable interaction model for the infectivities $\beta$. Numerical simulations show that the evolving dynamics of the network is effectively captured by a single scalar function of time, regardless of the
distribution of $\beta$ in the population.
The continuum limit of the network model allows a simple derivation of the simpler model, which is a single scalar delay differential equation (DDE), wherein the variation in $\beta$ appears
through an integral closely related to the moment generating function of $u=\sqrt{\beta}$. If the first few moments of $u$ exist, the governing DDE can be expanded in a series that shows a direct correspondence with the original compartmental DDE with a single $\beta$.
Even otherwise, the new scalar DDE can be solved using either numerical integration over $u$ at each time step, or with the analytical integral if
available in some useful form. Our work provides a new academic example of complete dimensional collapse, ties up an underlying continuum model for a pandemic with a simpler-seeming compartmental model, and will hopefully lead to new analysis of continuum models for epidemics.\\
\noindent {\bf Keywords:} pandemic, COVID-19, time delay, reduced order
\end{abstract}
\section{Introduction}
\label{intro}
The global pandemic of COVID-19 has prompted several studies of epidemic models from a dynamic systems point of view. 
Pandemics can be studied using simple mean-field models or compartmental models, where the entire population is divided into susceptible (S), exposed (E), infectious (I), quarantined (Q), and recovered (R) groups. More groups like hospitalized (H), vaccinated (V), etc., can be added, or groups can be removed, depending on modeling goals. These models are primarily developed based on the original SIR model due to Kermack and McKendrick~\cite{kermack1927contribution}. Many models include other complexities like prior immunity, temporary immunity transferred at birth, vaccination history, a carrier population that never recovers~\cite{hethcote2009basic},  reinfection due to loss of immunity after recovery~\cite{brauer2012mathematical}, exposed but asymptomatic populations~\cite{hethcote2000mathematics}, a quarantined population~\cite{gerberry2009seiqr}, and the influence of vital dynamics~\cite{hethcote1989three}.

The fidelity of such models can be improved by developing structured network models~\cite{zuzek2015epidemic,morita2016six,hasegawa2017efficiency,strona2018rapid,coelho2008epigrass,keeling2005networks}, where each node in the network is a compartmental model with different parameters. In other words, the entire population is divided into $N$ subgroups or structures. Each such subgroup or structure can have its own S, E, I, Q, and R subpopulations. These subgroups or structures can be based on age, lifestyle, demography, geography, or other aspects. The network topology~\cite{keeling2005networks} can be uniform, time-varying, or random. Recently, for example, the initial spread of COVID-19 in India was explored using an age-structured SIR network model with simulated social lockdown conditions~\cite{singh2020age}.

In contrast to such network models, an alternative high-dimensional approach uses reaction-diffusion \cite{barbera2013spread,ruan2007spatial,jones2009differential,schneckenreither2008modelling} type partial differential equations (PDEs) to capture the spatial and temporal evolution of an epidemic. Such models with spatial dimensions allow solutions with concentrated pockets of infection which emerge and spread. Sometimes, PDEs with a single spatial variable are converted into integro-differential equations~\cite{medlock2003spreading,ruan2007spatial}.

Time delays~\cite{van2002time} are often present in the progression of diseases due to latency and incubation times. Several researchers have studied the spread of infections like Zika, HIV, Hepatitis, and Influenza using delay differential equations ~\cite{young2019consequences,bocharov2000numerical,nelson2002mathematical,alexander2008delay,gourley2008dynamics,rakkiyappan2019}. Recently, an SEIQR~\cite{young2019consequences} model with time delays was proposed to study the progression of a generic epidemic. In recent work of our own~\cite{Vyas2020}, we have studied the time delayed SEIQR model of~\cite{young2019consequences} after neglecting loss of immunity over time, which is appropriate for a fast pandemic.

In the lumped approach of \cite{young2019consequences,Vyas2020}, the infectivity or transmission rate of the disease is modeled using a single positive parameter
\begin{equation} \label{pdef} \beta=\tilde{\beta} m.
\end{equation}
In Eq.~(\ref{pdef}), $\tilde {\beta}$ is a characteristic of the pathogen; and $m$, the density of contacts, is a characteristic of the behavior of individuals. 

In this paper, we address the situation where the parameter
$\beta$ is a distributed quantity over the entire population. After all, some people have greater exposure to infection than the population average (e.g., police personnel, people providing other essential services, hospital staff, as well as people less willing to cooperate with government-recommended social distancing measures). Others have less exposure than average (e.g., older people, people who cooperate more with recommended social distancing measures). Some have strong immune systems, and others have weak immune systems, which might be reflected in differing values of the underlying parameter $\tilde \beta$ in Eq.\ (\ref{pdef}). All these people interact to various degrees in modern society, where travel and mixing are common. The details of this variation, if incorporated, will yield a richer model that can hopefully make more realistic predictions. We clarify that we do not stratify the population by age, occupation, health status, or social behavior. In our model, all these factors contribute to a final effective $\beta$ for each person. We stratify the population according to
$\beta$ alone, because that is what affects the dynamics of infection.

We ask the following question: for a general distribution of $\beta$ in the population, and for a reasonable model for how different sections with different $\beta$'s interact during the pandemic, how does the continuum solution evolve from infinitesimal initial infection all the way to final saturation?

In the rest of this paper, we show the following. Under a simple {\em separable} interaction model for different $\beta$'s, and upon neglecting loss of immunity for those who have recovered from the disease, the continuum model reduces to two nonlinear integrodifferential equations with delays, with the continuously variable $\beta$ appearing as a parameter. However, due to a remarkable dimensional collapse, the dynamics is exactly described by a single scalar nonlinear delay differential equation without integrals, where the parameter $\beta$ appears {\em only} through the first derivative of its moment generating function. If some required moments of the distribution are finite, then a local expansion can be easily computed for small levels of overall infection, and the correpondence with the lumped model of \cite{young2019consequences,Vyas2020} is transparent and close. If the moments are unbounded, alternative, slightly more complex, single DDE approximations can still be developed, at least in principle. In this way, both the justification for the lumped model, as well as corrections needed for variable $\beta$, are exactly demonstrated. From a practical viewpoint, our results clarify the role of, e.g., higher versus lower variability of $\beta$ in the population in the spread, saturation, and containment of the disease. From an academic viewpoint, our work presents a new and satisfying example of extreme model order reduction. We mention that a preprint of this article has been uploaded at~\cite{Vyas20202} (\url{https://arxiv.org/abs/2004.12405}).
\section{Network Model}
In~\cite{young2019consequences,Vyas2020}, a single fixed parameter $\beta$ is used to model the whole population. Our presentation begins with a brief statement of this model. Figure~\ref{Schematic1} shows the schematic of the single-$\beta$ SEIQR model and is adapted from~\cite{young2019consequences}. The governing equations, included for completeness, are:
\begin{figure}[]
	\begin{center}
		\includegraphics[width=0.48\textwidth]{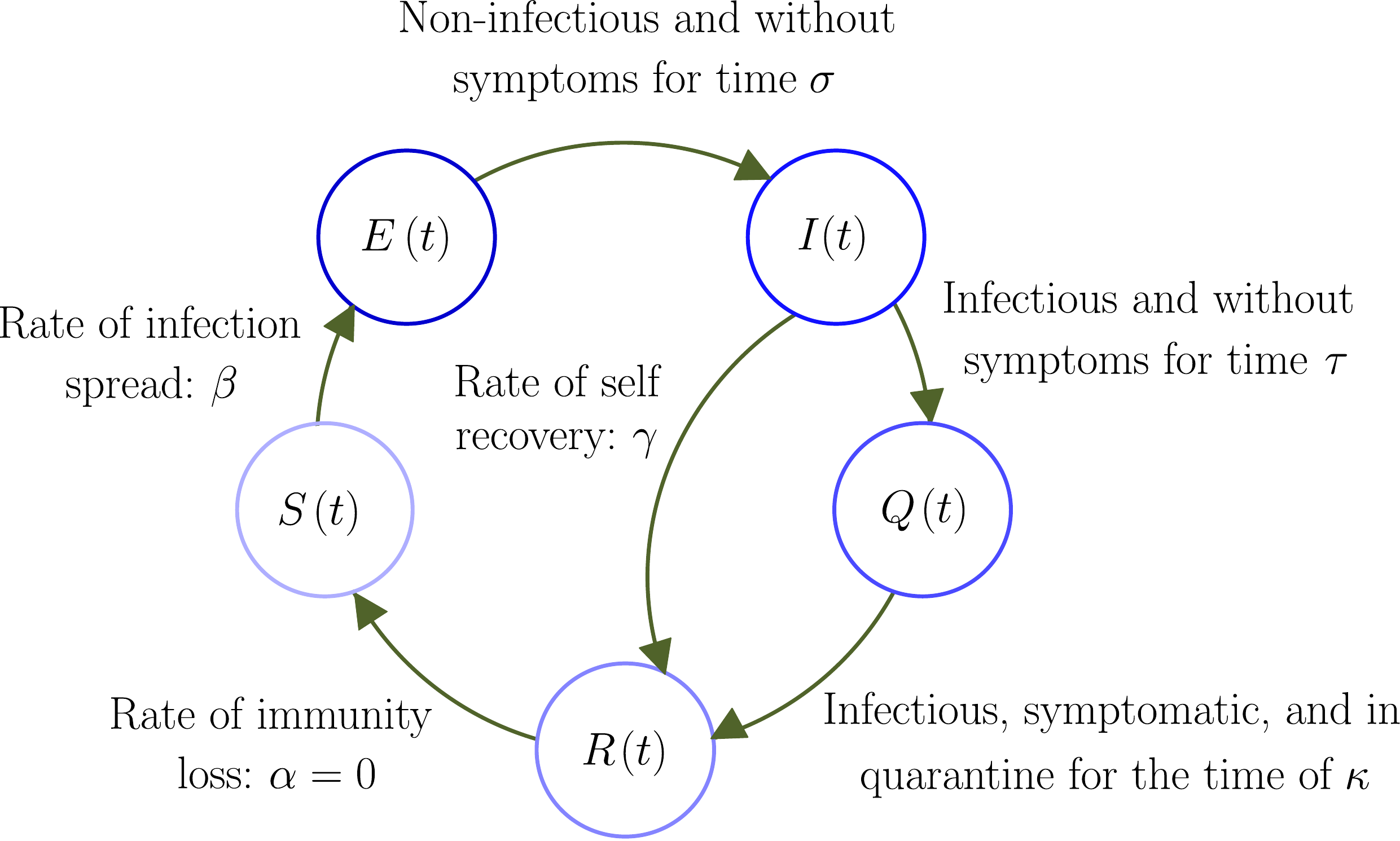}
	\end{center}
	\caption{The SEIQR compartmental model with delays.}
	\label{Schematic1}
\end{figure}
\begin{eqnarray}
\label{SdotR}
\dot{S}(t)&=&-\beta  S(t)I(t)+\alpha R(t)\\
\label{EdotR}
\dot{E}(t)&=&\beta \left[S(t)I(t)-S(t-\sigma)I(t-\sigma)\right]\\
\label{IdotR}
\dot{I}(t)&=& \beta  S(t-\sigma)I(t-\sigma)-\gamma I(t)\nonumber\\
&-&\beta  pe^{-\gamma\tau}S(t-\sigma-\tau)I(t-\sigma-\tau)\\
\label{QdotR}
\dot{Q}(t)&=&\beta  pe^{-\gamma\tau}S(t-\sigma-\tau)I(t-\sigma-\tau)\nonumber\\
&-&\beta pe^{-\gamma\tau}S(t-\sigma-\tau-\kappa)I(t-\sigma-\tau-\kappa)\\
\label{RdotR}
\dot{R}(t)&=&-\alpha R(t)+\gamma I(t)\nonumber\\
&+& \beta  pe^{-\gamma\tau} S(t-\sigma-\tau-\kappa)I(t-\sigma-\tau-\kappa)
\end{eqnarray}

 Here, $S(t)$  represents healthy individuals. The infection rate constant is $\beta$. Asymptomatic and infected individuals $E(t)$ remain non-infectious for $\sigma$ units of time. Later, they become infectious and are represented by $I(t)$, but show no symptoms for another $\tau$ units of time. When symptoms appear, these infected people are isolated or quarantined with probability $p$ for a time $\kappa$, and are represented by $Q(t)$.  A few asymptomatic but infectious individuals may recover on their own, at a rate $\gamma$. The cured population $R(t)$ after quarantine may lose immunity at a small rate $\alpha$, but we use $\alpha = 0$ over the time scale of a fast-spreading pandemic, like COVID-19.

\begin{figure}[htpb!]
	\begin{center}
		\includegraphics[width=0.45\textwidth]{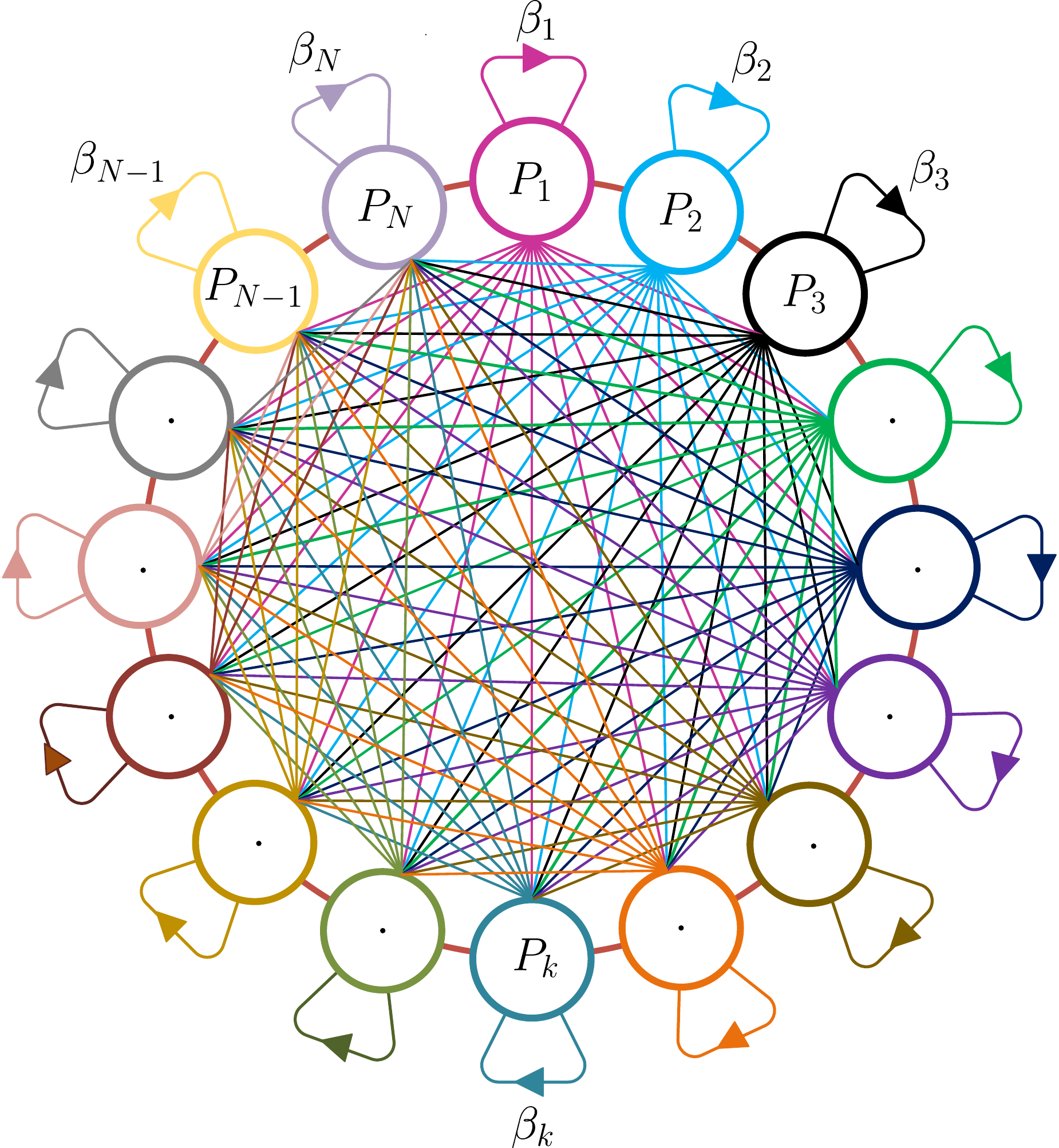}
	\end{center}
	\caption{A schematic representation of $N$ interacting population groups with different infection spread rates among each group. Every connection between two groups is bidirectional and symmetric, and every group is connected to all other groups (a dense network).}
	\label{Schematic2}
\end{figure}
In Eq.\ (\ref{SdotR}), the right hand side has two terms: the rate of change of the susceptible population $S(t)$ depends on an infection rate $\beta S(t) I(t)$ and a rate
of reinitroduction of susceptible people through loss of immunity modeled using $\alpha R(t)$. When $\beta$ varies within the population, we use
a network model. 
Figure~\ref{Schematic2} illustrates the idea. Here the total population is partitioned into $N$ sub groups $P_1$,$P_2$,...,$P_N$; and each group is modeled with different $\beta_k$ ranging from $\beta_1$ to $\beta_N$. The primary consequence is that in the network's equivalent of Eq.\ \ref{SdotR} for group $k$, the $I(t)$ must be replaced by something that represents interaction effects from all groups. We will model this using a quantity
$\Lambda(t)$ that applies to the entire network, as explained below.

The main idea is simple. A susceptible person in group $k$, for any $1 \le k \le N$, can get infected through interaction with a person in any group
$r$, with $1 \le r \le N$. Thus, the total infection rate of the single-$\beta$ model, i.e., $\beta S(t) I(t)$, is to be replaced by the sum of contributions from {\em all} groups. The contribution from one single group $r$ is taken to be
$$F(\beta_k,\beta_r) S_k(t) I_r(t),$$
for some suitable interaction coefficient $F(\beta_k,\beta_r)$ that has to be specified as part of the mathematical model.
The function $F(\beta_k,\beta_r)$ is understood to be symmetric~\cite{blackwood2018introduction}, i.e.,
$$F(x,y) = F(y,x).$$
It must monotonically increase with respect to each of its arguments, i.e.,
$$\frac{\partial F(x,y)}{\partial x} > 0, \quad \frac{\partial F(x,y)}{\partial y} > 0.$$
Clearly, we require
$$F(x,0)=0$$
regardless of $x$, because a person who meets nobody at all will not get infected.
Further, it seems reasonable to assume that if a person who mixes very little meets another person who mixes very little, then the probability of infection spreading is much lower than if the second person mixes widely. Mathematically, we assume that
\begin{equation}
\label{sep}
\mbox{if}\, 0 < \beta_k \ll \beta_r, \, \mbox{ then } \, F(\beta_k, \beta_k) \ll F(\beta_k, \beta_r) \ll F(\beta_r, \beta_r).
\end{equation}
All the above conditions are met by the {\em separable} choice
\begin{equation} 
\label{fd0} 
F(\beta_k,\beta_r) = \sqrt{\beta_k} \sqrt{\beta_r}. 
\end{equation}
here could be other formulas for $F(\beta_k,\beta_r)$ satisfying the conditions mentioned above, but the separable nature of Eq.~(\ref{fd0}) makes it simple as well as appealing.

Finally, in the special case where the entire population moderates its behavior to produce a single effective $\beta$, the net effect should ideally reduce to
Eq.\ (\ref{SdotR}), and this is something that will turn out to be true for Eq.\ (\ref{fd0}).

The above choice of $F(\beta_k,\beta_r)$ leads us to define
\begin{equation}
\Lambda(t)=\sum_{r=1}^{N}\sqrt{\beta_{r}}I_{r}(t),
\label{Lambda}
\end{equation}
and the dynamics of the network is governed by
\begin{eqnarray}
\label{Sdot1}
\dot{S}_{k}(t)&=&-\sqrt{\beta_{k}}S_{k}(t)\Lambda(t)+\alpha R_{k}(t)\\
\label{Edot1}
\dot{E}_{k}(t)&=&\sqrt{\beta_{k}}\left[S_{k}(t)\Lambda(t)-S_{k}(t-\sigma)\Lambda(t-\sigma)\right]\\
\label{Idot1}
\dot{I}_{k}(t)&=&\sqrt{\beta_{k}}S_{k}(t-\sigma)\Lambda(t-\sigma)-\gamma I_k(t)\nonumber\\
&-&\sqrt{\beta_{k}}pe^{-\gamma\tau}S_{k}(t-\sigma-\tau)\Lambda(t-\sigma-\tau)\\
\label{Qdot1}
\dot{Q}_k(t)&=&\sqrt{\beta_{k}} pe^{-\gamma\tau}S_k(t-\sigma-\tau)\Lambda(t-\sigma-\tau)\nonumber\\ &-&\sqrt{\beta_{k}}pe^{-\gamma\tau}S_k(t-\sigma-\tau-\kappa)\Lambda(t-\sigma-\tau-\kappa)\\
\label{Rdot1}
\dot{R}_k(t)&=&-\alpha R_k(t)+\gamma I_k(t)\nonumber\\
&+&\beta_{k} pe^{-\gamma\tau}S_k(t-\sigma-\tau-\kappa)\Lambda(t-\sigma-\tau-\kappa)\\
k&=&1,2,...,N\nonumber
\end{eqnarray}
In this model, except for $\beta_k$, all other parameters namely $\sigma$, $\tau$, $\kappa$, $p$, $\gamma$, and $\alpha$ are the same for each node in the network. This assumption is motivated by the idea that, in a well-functioning society, the degree of mixing practiced by individuals varies a lot more than the time taken for an infected person to be detected and quarantined; and that the rate of self-recovery $\gamma$ is a biological quantity independent of an individual's social behavior. Note that we assume the loss of immunity rate, $\alpha$, to be zero.

We observe from Eqs.~(\ref{Sdot1})-(\ref{Rdot1}) that if $\alpha=0$ the states $E_k(t)$, $Q_k(t)$, and $R_k(t)$ become slave variables. Only $S_k(t)$ and $I_k(t)$ need to be solved for. Further, we can always scale time by setting $\sigma=1$, making $\tau$, $\beta_{k}$, and $\gamma$ effectively dimensionless. Equations~(\ref{Sdot1})-(\ref{Rdot1}), after dropping $\dot{E}$, $\dot{Q}$ and $\dot{R}$ and defining
$$\nu=1+\tau$$
become:
\begin{eqnarray}
\label{Sdot2}
\dot{S}_{k}(t)&=&-\sqrt{\beta_{k}}S_{k}(t)\Lambda(t)\\
\label{Idot2}
\dot{I}_{k}(t)&=&\sqrt{\beta_{k}}S_{k}(t-1)\Lambda(t-1)-\gamma I_k(t)\nonumber\\
&-&\sqrt{\beta_{k}}pe^{-\gamma\tau}S_{k}(t-\nu)\Lambda(t-\nu)\\
k&=&1,2,...,N.\nonumber
\end{eqnarray}

\section{Initial Numerical Observations}
We have integrated Eqs.~(\ref{Sdot2}) and (\ref{Idot2}) using Matlab's built-in solver {\tt dde23} for many different initial functions or history functions, with error tolerances set to be $10^{-7}$ or better.  For ease of presenting results, we define 
	$$\hat{\beta}=[\beta_{1},\beta_{2},...\beta_{N}]^{T},$$
	$$\hat{S}(t)=\left[S_{1}(t),\thinspace S_{2}(t),\thinspace...,S_N(t)\right]^{T},$$
	$$\hat{I}(t)=\left[I_{1}(t),\thinspace I_{2}(t),\thinspace...,I_N(t)\right]^{T}.$$
	It should be noted that $\beta_1<\beta_2<\dots<\beta_N$. The history functions for numerical integration are selected in terms of nonnegative functions $\hat U(\hat \beta)$ and $\hat V(\hat \beta)$ as
\begin{eqnarray}
\label{Sinit}
\hat{S}(t)&=&\hat U(\hat \beta)-10^{-8}\left(1+\frac{t}{\nu}\right)
,-\nu\le t\le 0,\\
\label{Iinit}
\hat{I}(t)&=&\hat V(\hat \beta)\times 10^{-8}\left(1+\frac{t}{\nu}\right)
,-\nu\le t\le 0.
\end{eqnarray}
In the above, $\hat U(\hat \beta)$ was chosen to satisfy $||\hat{S}(-\nu)||_{1}=1$. It is clear that $||\hat{I}(-\nu)||_{1}=0$
regardless of $\hat V(\hat \beta)$.
Such initial conditions, with small initial infection, are of interest because we wish to study whether the infection remains small or grows significantly within the population. It is clear that $\hat U(\hat \beta)$ plays the role of a probability density function of people's $\hat \beta$-values in the population.
If $\hat U(\hat \beta)$ is large for small values of $\hat \beta$ and decays rapidly for large values of $\hat \beta$, then most people are in a regime of low infection rates. Conversely, if $\hat U(\hat \beta)$ decays slowly with increasing values of $\hat \beta$, then a signifcant proportion of the population is in a regime of high
infection rates. The dynamic consequences of such different distributions will be studied using initial numerical simulations in this section, before proceeding to analytical treatment.

The net effect of the pandemic, or net damage, is taken to be
$$\hat{D}(t)=\hat{S}(-\nu)-\hat{S}(t).$$
For total percentage affected, we use
$$D(t)=||\hat{D}(t)||_{1}\times100=\left[1-||\hat{S}(t)||_{1}\right]\times100.$$

\begin{figure*}[htpb!]
	\begin{center}
		\subfigure[]{\includegraphics[width=0.45\textwidth]{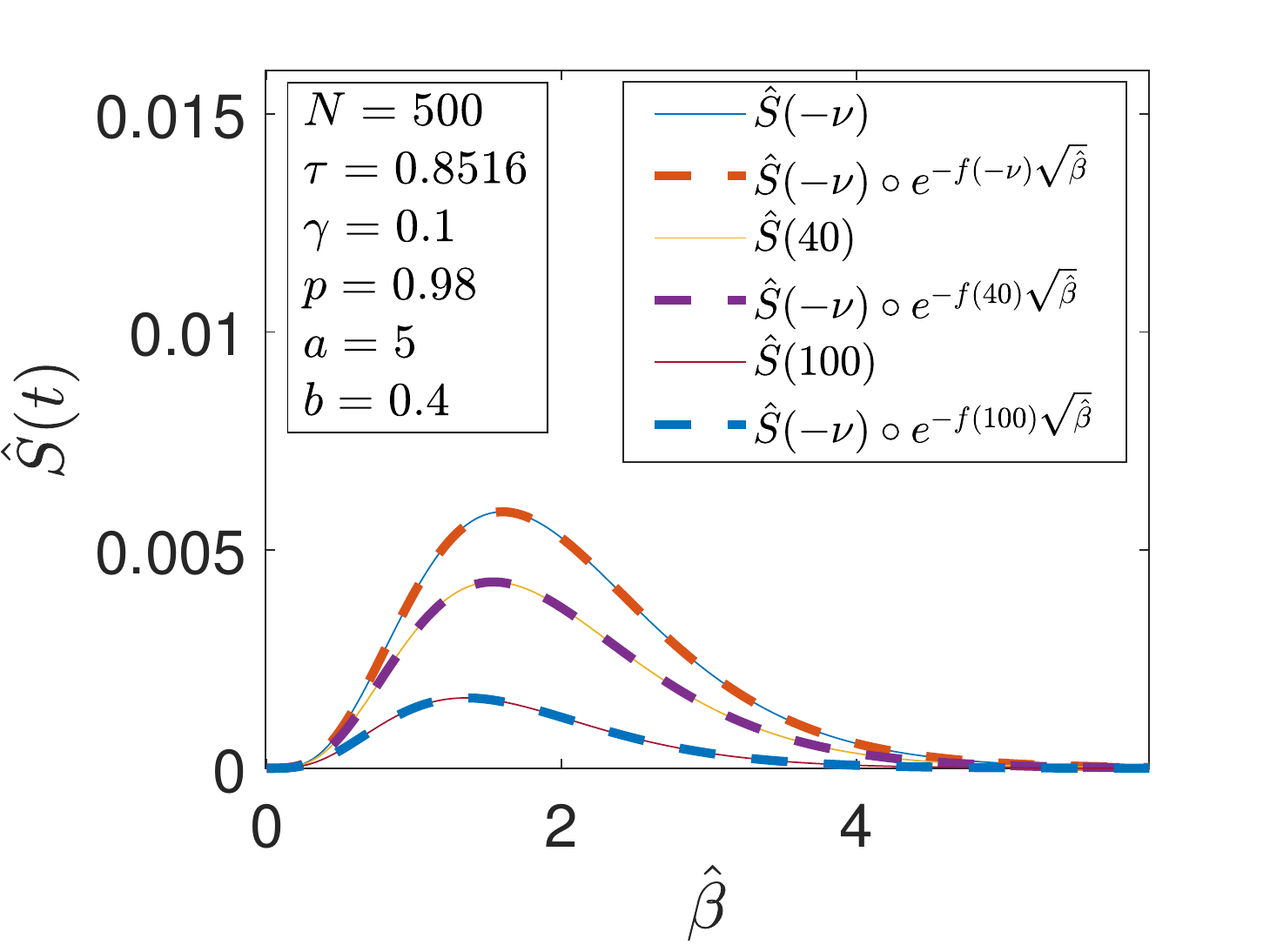}\label{case1a}}
		\subfigure[]{\includegraphics[width=0.45\textwidth]{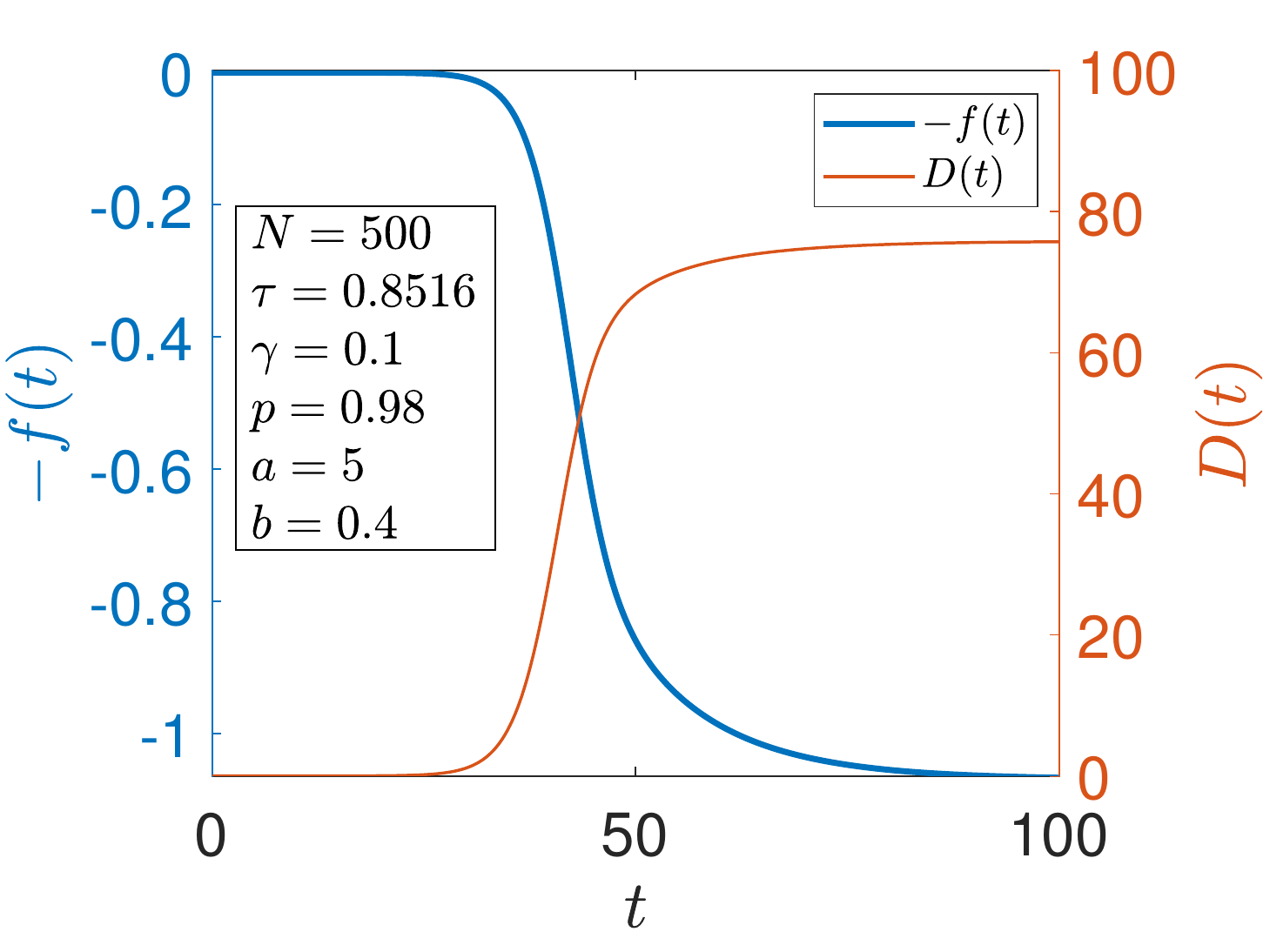}\label{case1b}}
		\subfigure[]{\includegraphics[width=0.45\textwidth]{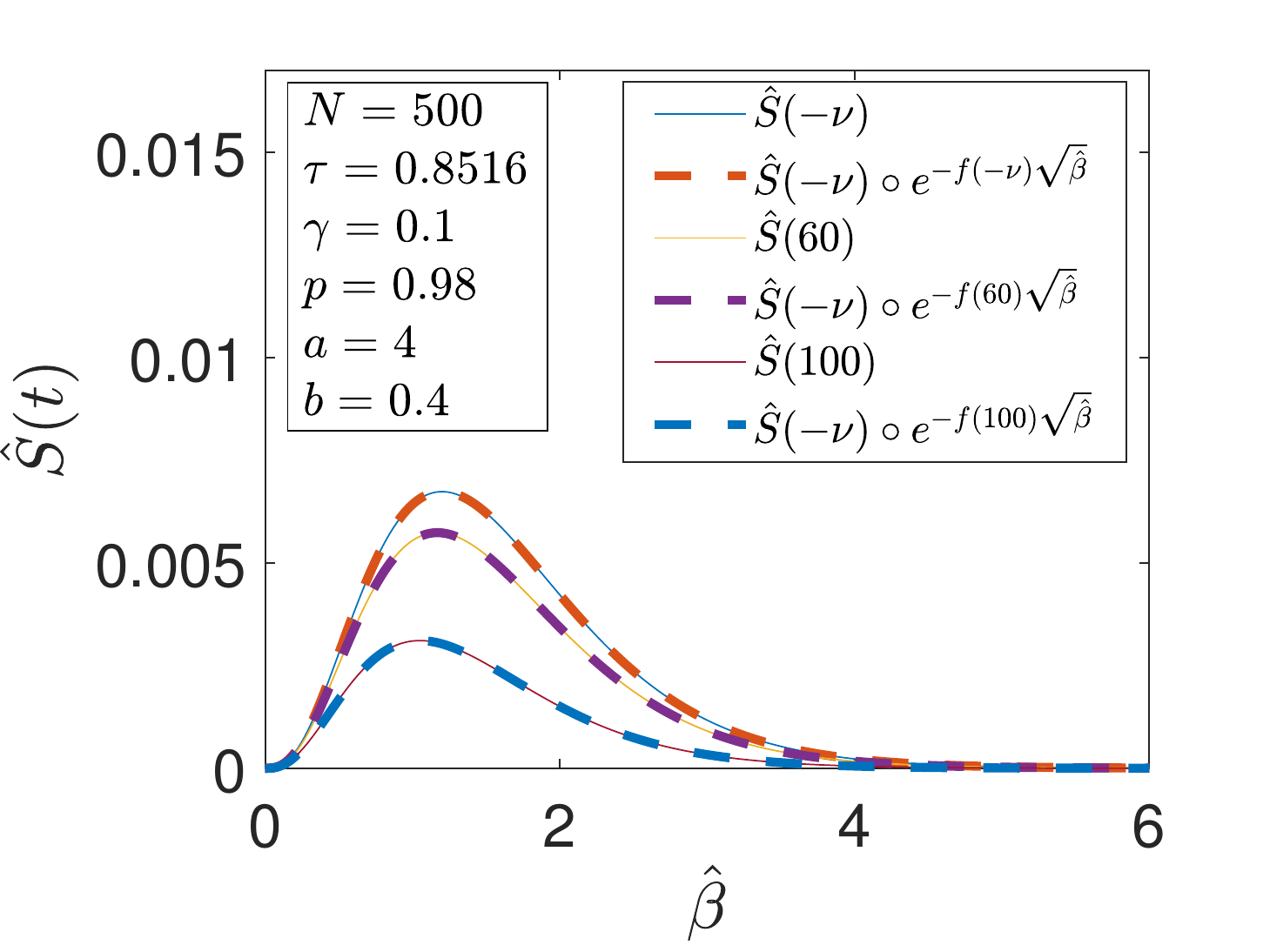}\label{case1c}}
		\subfigure[]{\includegraphics[width=0.45\textwidth]{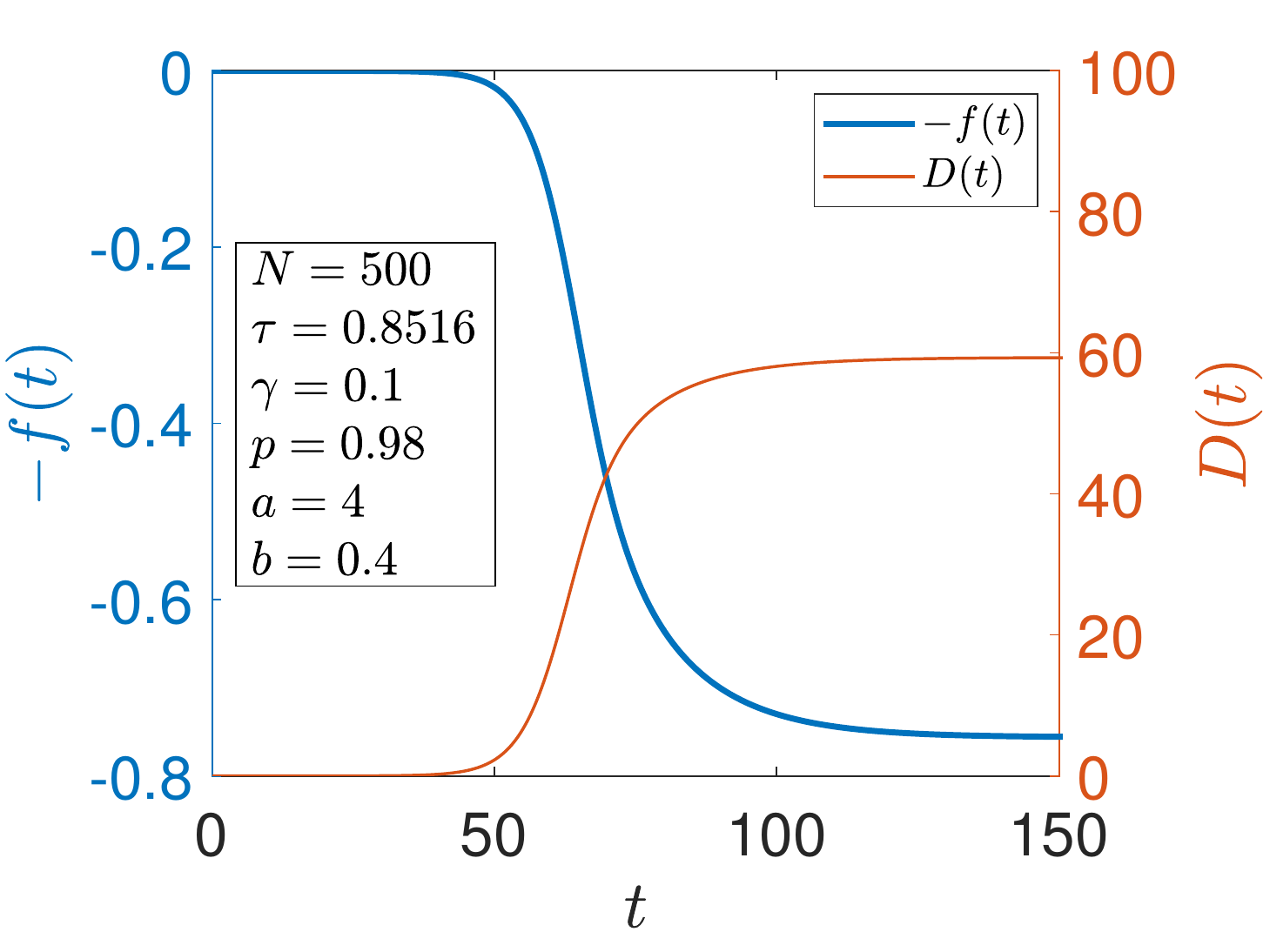}\label{case1d}}
		\subfigure[]{\includegraphics[width=0.45\textwidth]{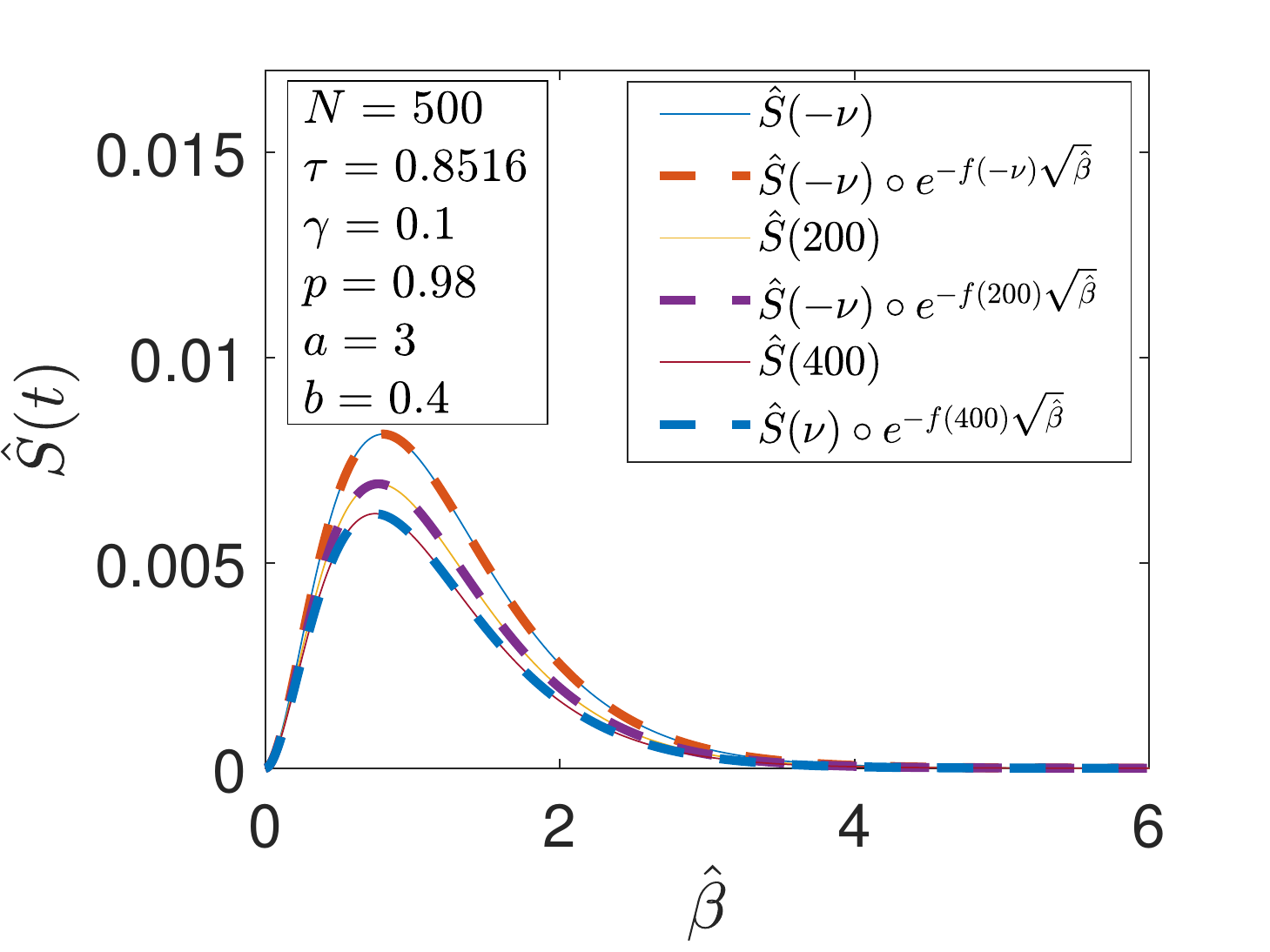}\label{case1e}}
		\subfigure[]{\includegraphics[width=0.45\textwidth]{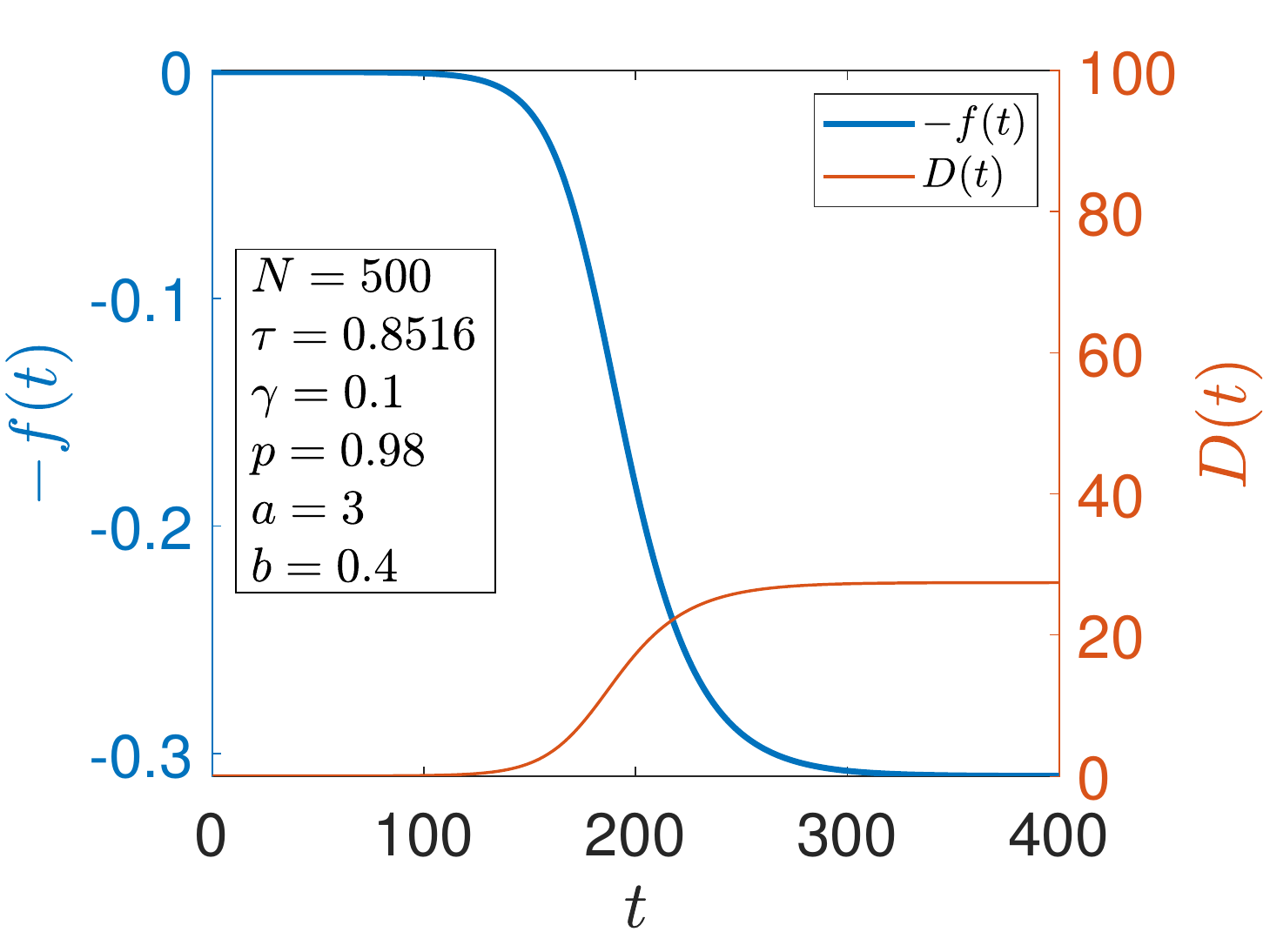}\label{case1f}}
	\end{center}
	\caption{Simulations with different initial values $\hat{S}(-\nu)$. Here, a Gamma distribution is used for $\hat{U}(\hat{\beta})=\frac{1}{b^{a}\Gamma(a)}\hat{\beta}^{(a-1)}e^{\frac{-\hat{\beta}}{b}}$, along
		with $\hat{V}(\hat{\beta})=\hat{U}(\hat{\beta})$. Also shown is the fit of the form of Eq.\ (\ref{fithai}) at different instants of time. We note that as the
		$\beta$ values in the distribution get smaller on average, $D(\infty)$ decreases.
	}
	\label{Figure2}
\end{figure*}

For our initial numerical case study, we use $N=500$ and select $\hat{\beta}$ values uniformly spaced between $\beta_1=0$ and $\beta_N=6$, i.e., $\hat{\beta}$ values are distributed within a strictly finite range.
Figure~\ref{Figure2}(a) shows $\hat{S}(t)$ at three different time instants at $t=-\nu$ (initial time), $t=40$, and $t=100$. Also shown is a fit of the form 
\begin{equation} \label{fithai}
\hat S(t) = \hat S( -\nu) \circ e^{-f(t) \sqrt{\hat{\beta}}},
\end{equation}
at these three instants of time. Where ``$\circ$'' denotes elementwise multiplication of two arrays. The initial conditions were
$$\hat{U}(\hat{\beta})=\frac{1}{b^{a}\Gamma(a)}\hat{\beta}^{(a-1)}e^{\frac{-\hat{\beta}}{b}},$$
which matches the Gamma distribution, with $a=5$ and $b=0.4$ (technically, the Gamma distribution has infinite support, but for the parameters chosen it has decayed to tiny values for $\hat{\beta}=6$). Also, the other initial function $\hat V(\hat \beta)$ was taken to be identical to
$\hat U(\hat \beta)$ in the simulations of Fig.~\ref{Figure2}.

We emphasize that in Fig~\ref{Figure2}(a), $f(-\nu)=0$ by definition and there are only {\em two} fitted
numbers: $f(40)$ and $f(100)$. The match is essentially perfect, and shows that the evolution of all the $S_k(t)$ together have a one-dimensional behavior. The solution at any time $t$ is fitted, essentially
perfectly, by a function of the form shown in Eq.~(\ref{fithai}).

Figure~\ref{Figure2}(b) shows $-f(t)$ and the total infected population $D(t)$ during the progression of the pandemic. Parameters used for generating the results are shown in the text boxes inside the figure.

Other simulations with different $a$ and $b$ in the underlying Gamma distribution are shown in subplot pairs
Figs~\ref{Figure2}(c)-(d) and Figs~\ref{Figure2}(e)-(f). The quality of the match remains excellent, as is shown graphically for different intermediate time instants in Figs~\ref{Figure2}(c) and (e), although the net percentage of people infected changes with $a$ and $b$ (more people are infected if typical $\hat{\beta}$ values in the distribution are higher).

\begin{figure*}[htpb!]
	\begin{center}
		\subfigure[]{\includegraphics[width=0.45\textwidth]{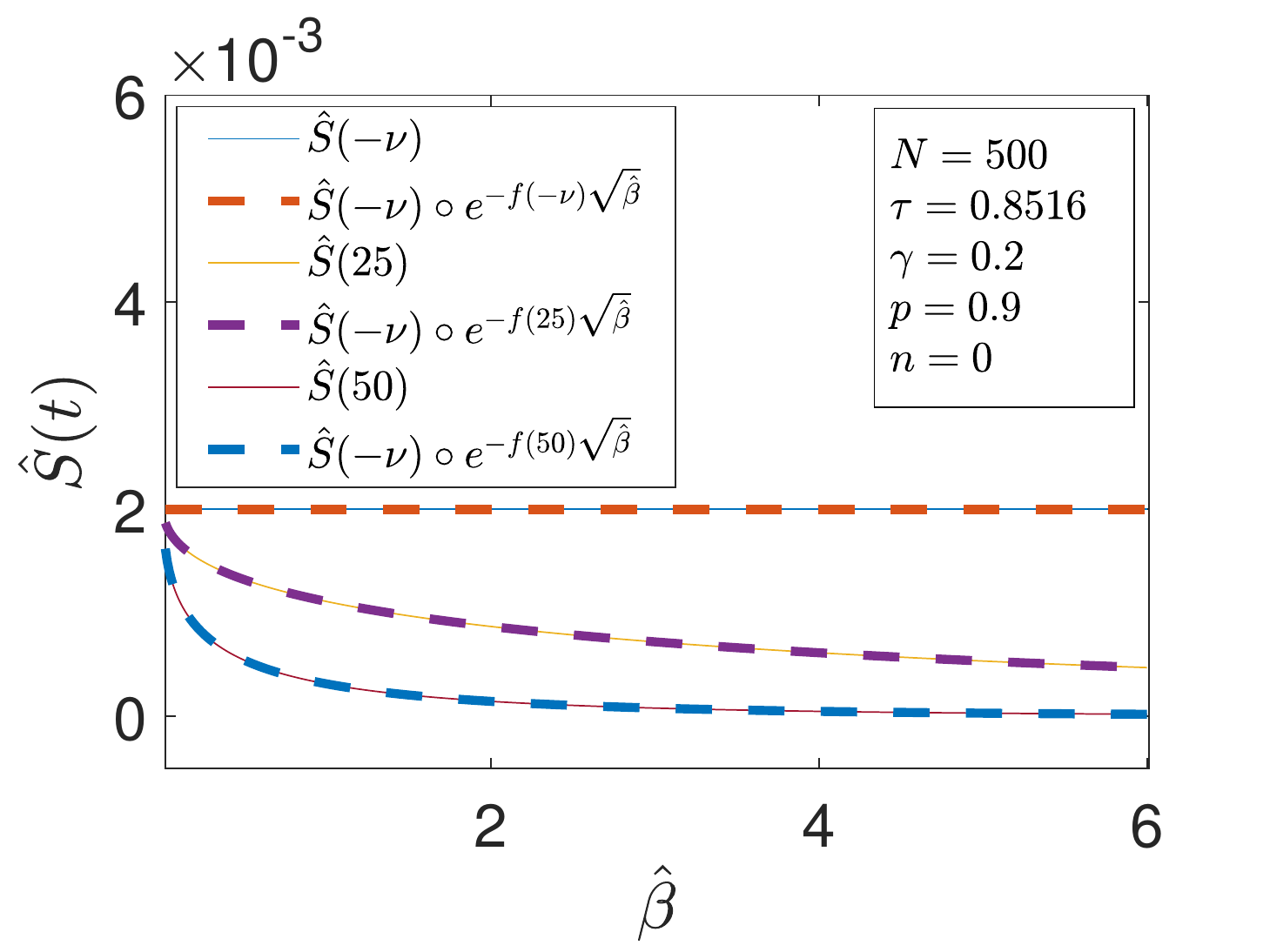}\label{case2a}}
		\subfigure[]{\includegraphics[width=0.45\textwidth]{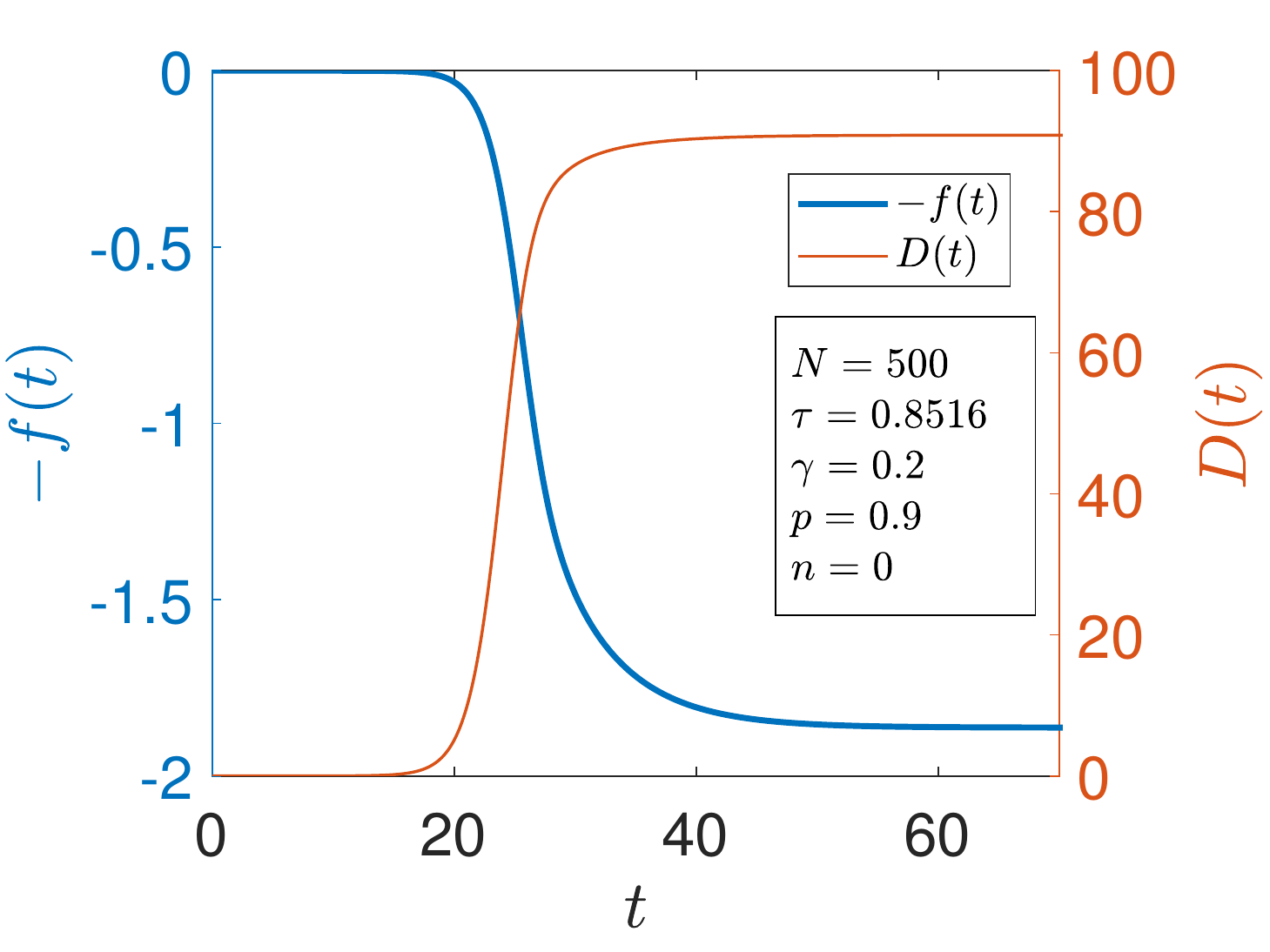}\label{case2b}}
		\subfigure[]{\includegraphics[width=0.45\textwidth]{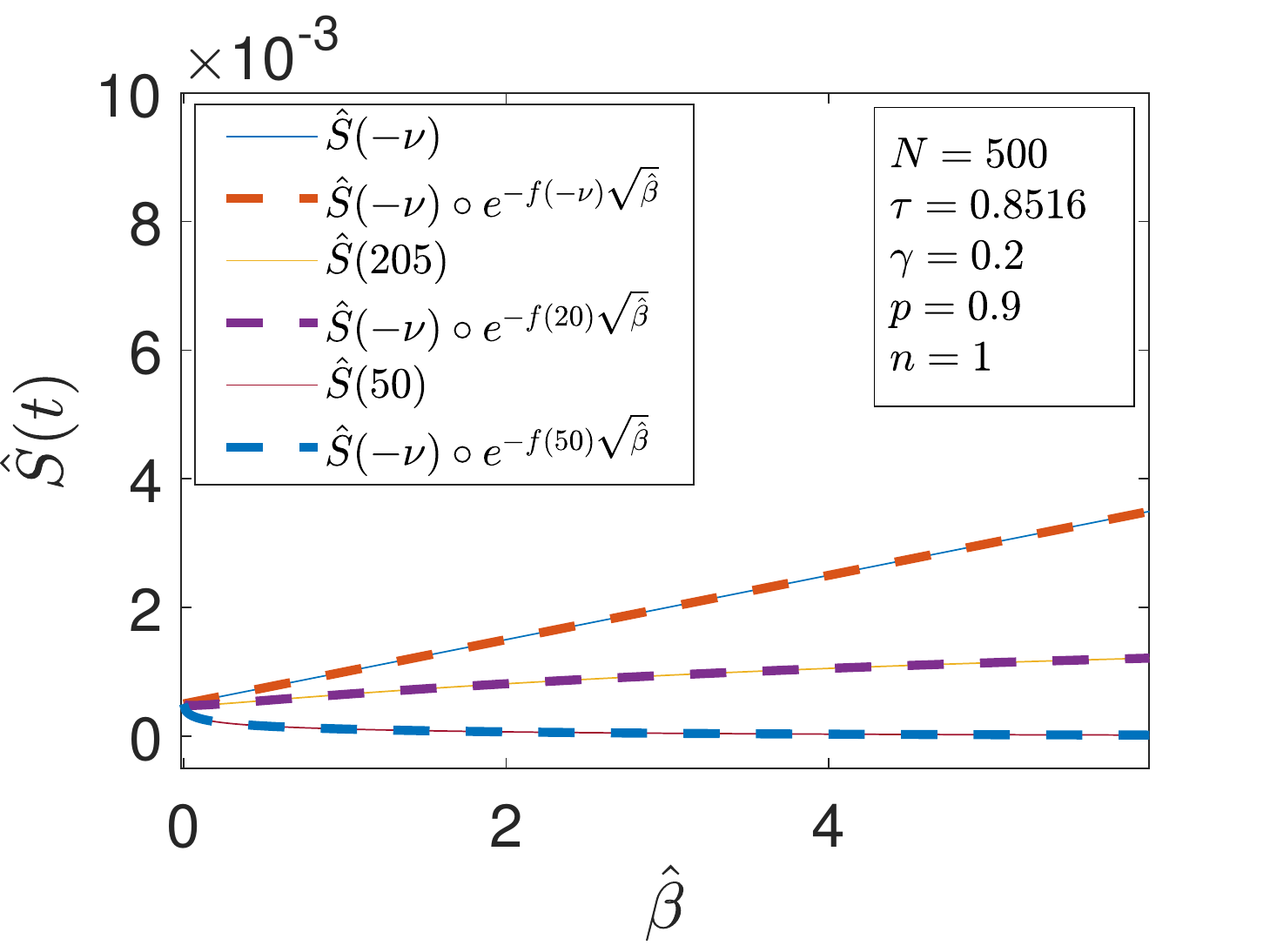}\label{case2c}}
		\subfigure[]{\includegraphics[width=0.45\textwidth]{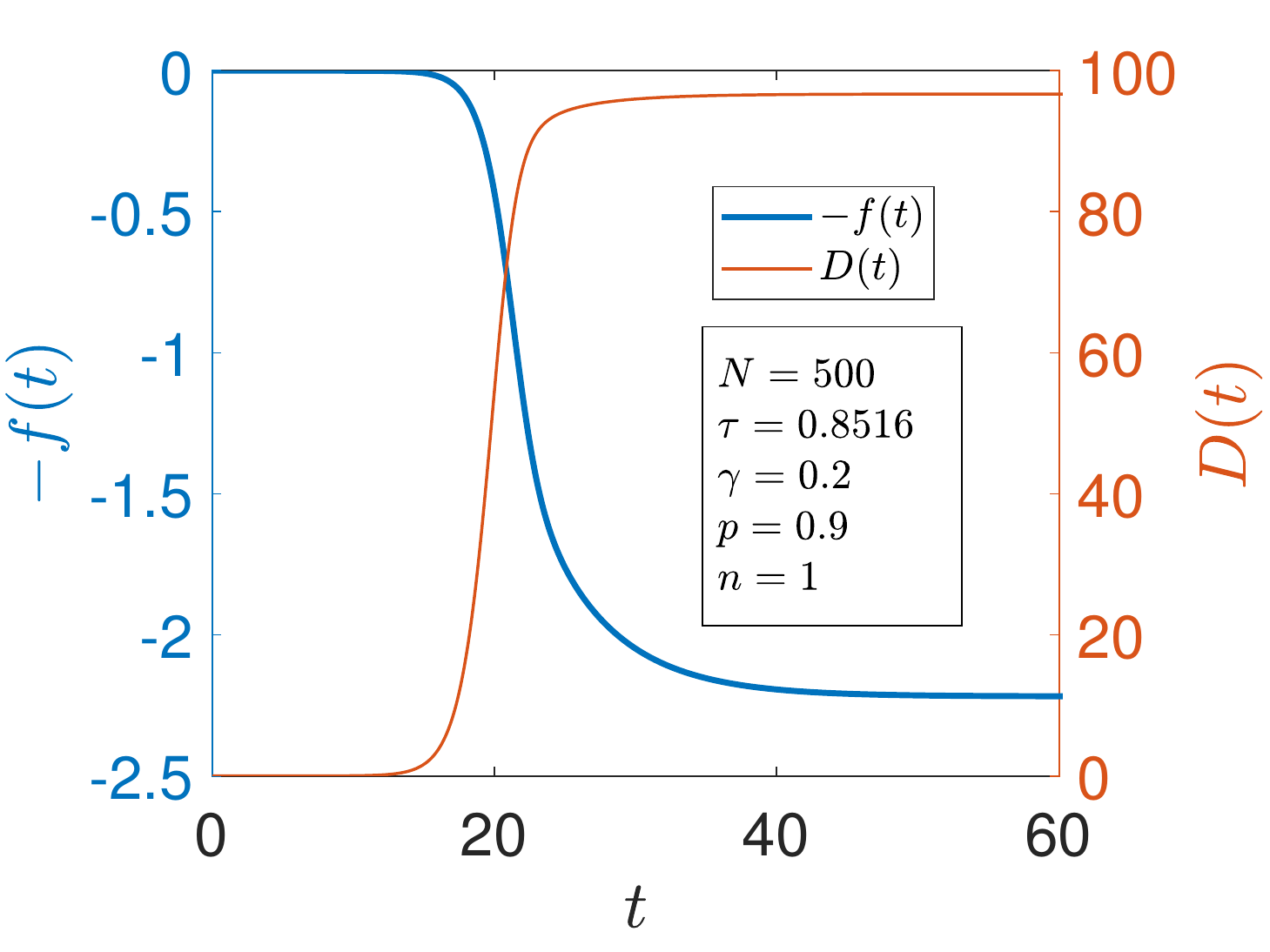}\label{case2d}}
		\subfigure[]{\includegraphics[width=0.45\textwidth]{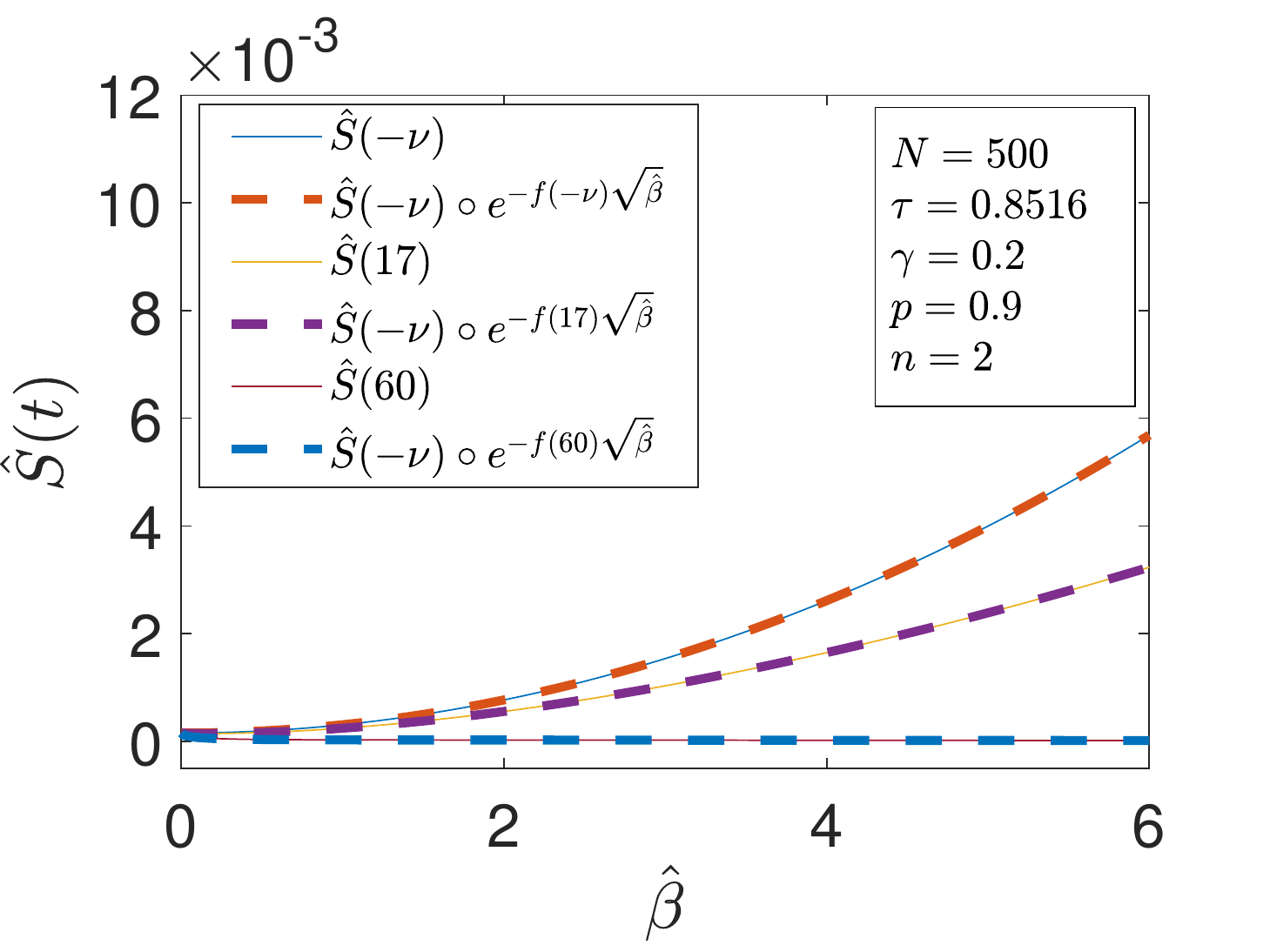}\label{case2e}}
		\subfigure[]{\includegraphics[width=0.45\textwidth]{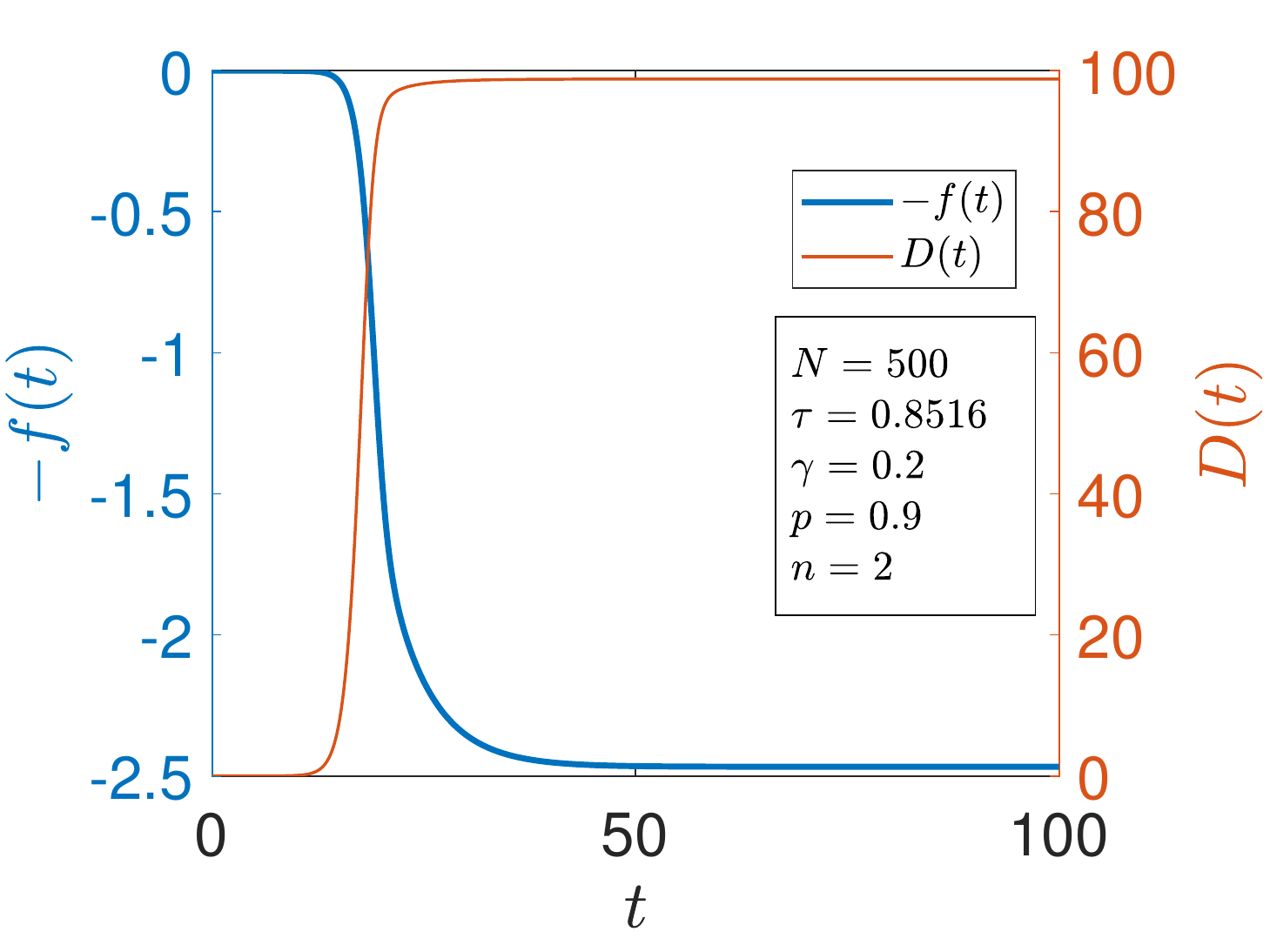}\label{case2f}}
	\end{center}
	\caption{Simulations with different initial values  $\hat{S}(-\nu)=\hat{U}(\hat{\beta})$. Here, finitely supported polynomials of the form $\hat{U}(\hat{\beta})=r_n \cdot \left ( 1+\hat{\beta}^n \right ),\,0\le \hat{\beta} \le 6$ are considered, with $r_n$ chosen to make the sum (or 1-norm) of $\hat{U}(\hat{\beta})$ equal to 1. The distribution of initial values for $\hat{I}(\hat{\beta})$ uses
		$\hat{V}(\hat{\beta})$ that is random, initially uniformly distributed between 1 and 2, and then normalized to unit 1-norm. Also shown is the fit
		$\hat{S}(t)=\hat{S}(-\nu)\circ e^{-f(t)\sqrt{\hat{\beta}}}$ at different instants of time. These results demonstrate that provided the intial infected population distribution is small, the subsequent evolution obeys a simple one-dimensional description with a scalar variable $f(t)$.
	}
	\label{Figure3}
\end{figure*}

In Fig~\ref{Figure3}, we show results for initial conditions $\hat{U}(\hat{\beta})$ chosen arbitrarily on the finite interval $[0,6]$, without an underlying asymptotic approach to zero for large $\hat{\beta}$ (see figure caption). The choices for $\hat{U}(\hat{\beta})$ are now polynomials on the interval
$[0,6]$. From Fig~\ref{Figure3}, we observe that for several different $\hat{U}(\hat{\beta})$ and $\hat{V}(\hat{\beta})$ and for generic small-infection history functions (Eq.~(\ref{Sinit}) and Eq.~(\ref{Iinit})), a fit of the form given by  Eq.~(\ref{fithai}) is essentially exact. In other words, the variation over $\hat{\beta}$ is one-dimensional, in terms of a scalar $f(t)$. We refer to this great reduction in dimensionality, where the variation with respect to the continuous variable $\hat{\beta}$ is accounted for by a single scalar, as {\em complete dimensional
	collapse}. Numerics indicate the collapse is exact.

It remains only to extract the governing equation for the scalar $f(t)$, and we will do this using a continuum formulation.

\section{Continuum Model}
We consider a continuum limit of Eqs.~(\ref{Sdot2}) and (\ref{Idot2}), as $N\rightarrow\infty$. We replace the summation in Eq.~(\ref{Lambda}) with integrals and assume $S(\beta,t)$ and $I(\beta,t)$ to be functions of $\beta$ and $t$. In the continuum limit, Eqs.~(\ref{Sdot2}) and (\ref{Idot2}) become
\begin{eqnarray} 
\label{e3}
\dot S(\beta,t) &=& - \sqrt{\beta} S(\beta,t) \int_0^{\infty} \sqrt{\xi} I(\xi,t) \, d \xi\\
\label{e4}
\dot I(\beta,t) &=& \sqrt{\beta} S(\beta,t-1) \int_0^{\infty} \sqrt{\xi} I(\xi,t-1) \, d \xi- \gamma I(\beta,t) \nonumber\\
&-&\bar p \sqrt{\beta} S(\beta,t-\nu) \int_0^{\infty} \sqrt{\xi} I(\xi,t-\nu) \, d \xi ,
\end{eqnarray}
where
\begin{equation} \label{pbareq} \bar p = p e^{-\gamma \tau}. \end{equation}
In the above integrals, if $\beta$ takes only finite values (in the numerical examples above, $\beta$ was between 0 and 6), then the upper limits of the integrals can be replaced with finite values; we write $\infty$ as a formal upper limit. Note that Eqs.\ (\ref{e3}) and (\ref{e4}) represent infinitely many coupled delay differential equations, parameterized by the distributed infectivity parameter $\beta$.

The fraction of the population that is susceptible is now understood to be
$$\int_{0}^{\infty} S(\beta,t) \, d \beta,$$
and the fraction of the presently infectious population is understood to be
$$\int_{0}^{\infty} I(\beta,t) \, d \beta.$$
At the start of the pandemic, we have initial conditions
that satisfy
$$\int_{0}^{\infty} S(\beta,0) \, d \beta \approx 1,$$
and
$$\int_{0}^{\infty} I(\beta,0) \, d \beta \ll 1,$$
by which we mean a tiny initiation of infection. 

Introduction of a tiny infected population can, with suitable initial distributions of $S$, make the pandemic grow. A society that has strong social distancing will have $S(\beta,0)$ decaying to zero rapidly  with increasing $\beta$, and may not see strong growth of the infection. Conversely, a society where the population has greater social mixing, i.e., $S(\beta,0)$ that decays slowly with increasing $\beta$, may see an outbreak of the infection.

A point to note during numerical solution of Eqs.~(\ref{e3}) and (\ref{e4}) is that $S(\beta,t)$ and $I(\beta,t)$ are nonnegative. It is in principle possible for $I(\beta,t)$ to start positive and become exactly zero at some instant (based on hypothetical initial functions used in the delayed variables), and it corresponds to all infected people becoming quarantined before fresh people are infected. However, such a situation corresponds to the pandemic being quenched by eliminating infection, and is not of interest when either infection cannot be eliminated, or when even elimination of infection leaves the system unstable (i.e., introduction of infinitesimal infection leads to an outbreak). As seen in the numerical solutions of Figs~\ref{Figure2} and \ref{Figure3}, there are clearly also other solutions of interest where the pandemic progresses, infects a percentage of the population, and reaches an eventual equilibrium with the infection not progressing further. For such solutions, the nonnegativity constraint of $S(\beta,t)$ and $I(\beta,t)$ turns out to be satisfied.

\begin{figure*}[htpb!]
	\begin{center}
		\subfigure[]{\includegraphics[width=0.45\textwidth]{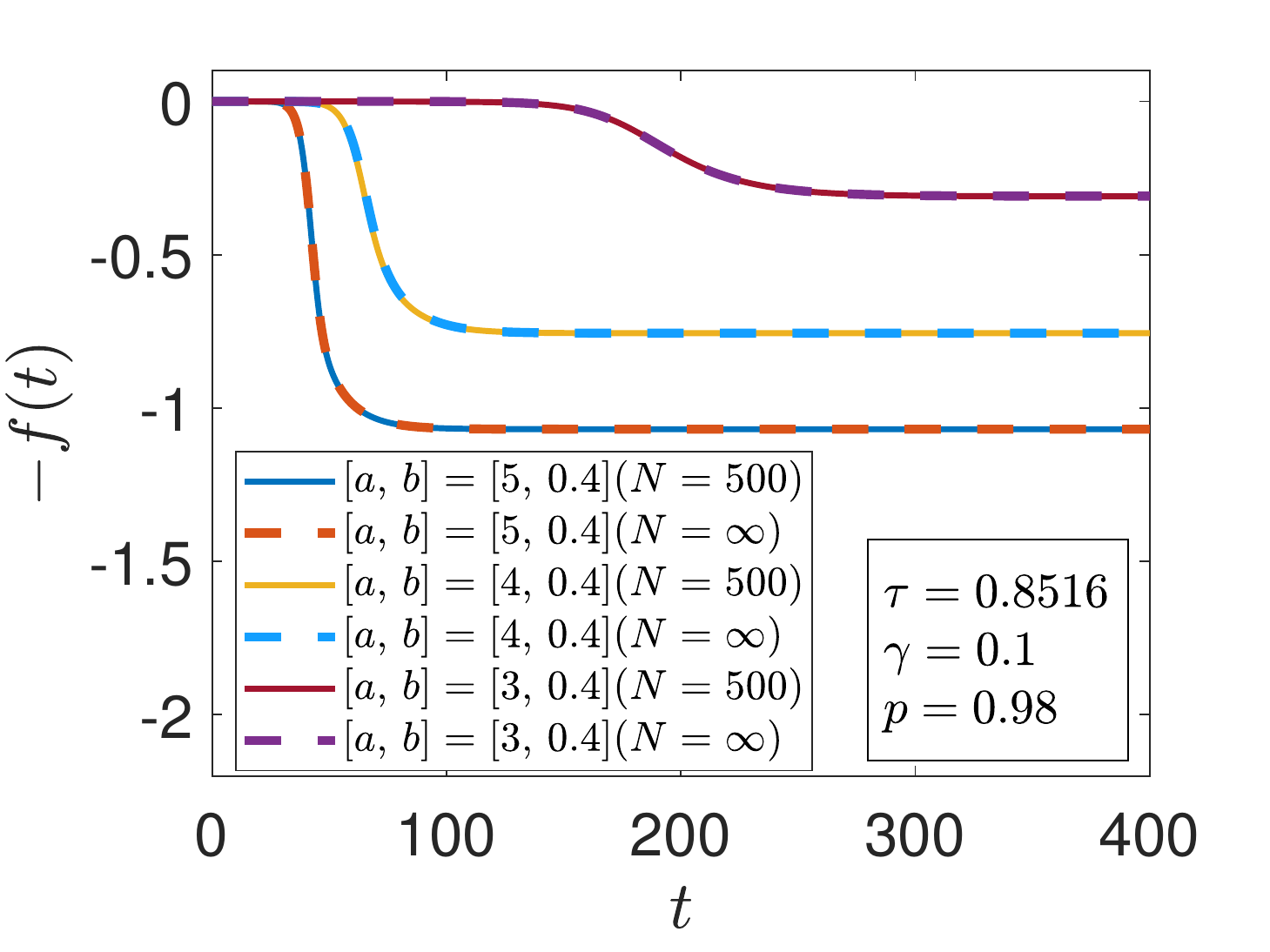}\label{case3a}}
		\subfigure[]{\includegraphics[width=0.45\textwidth]{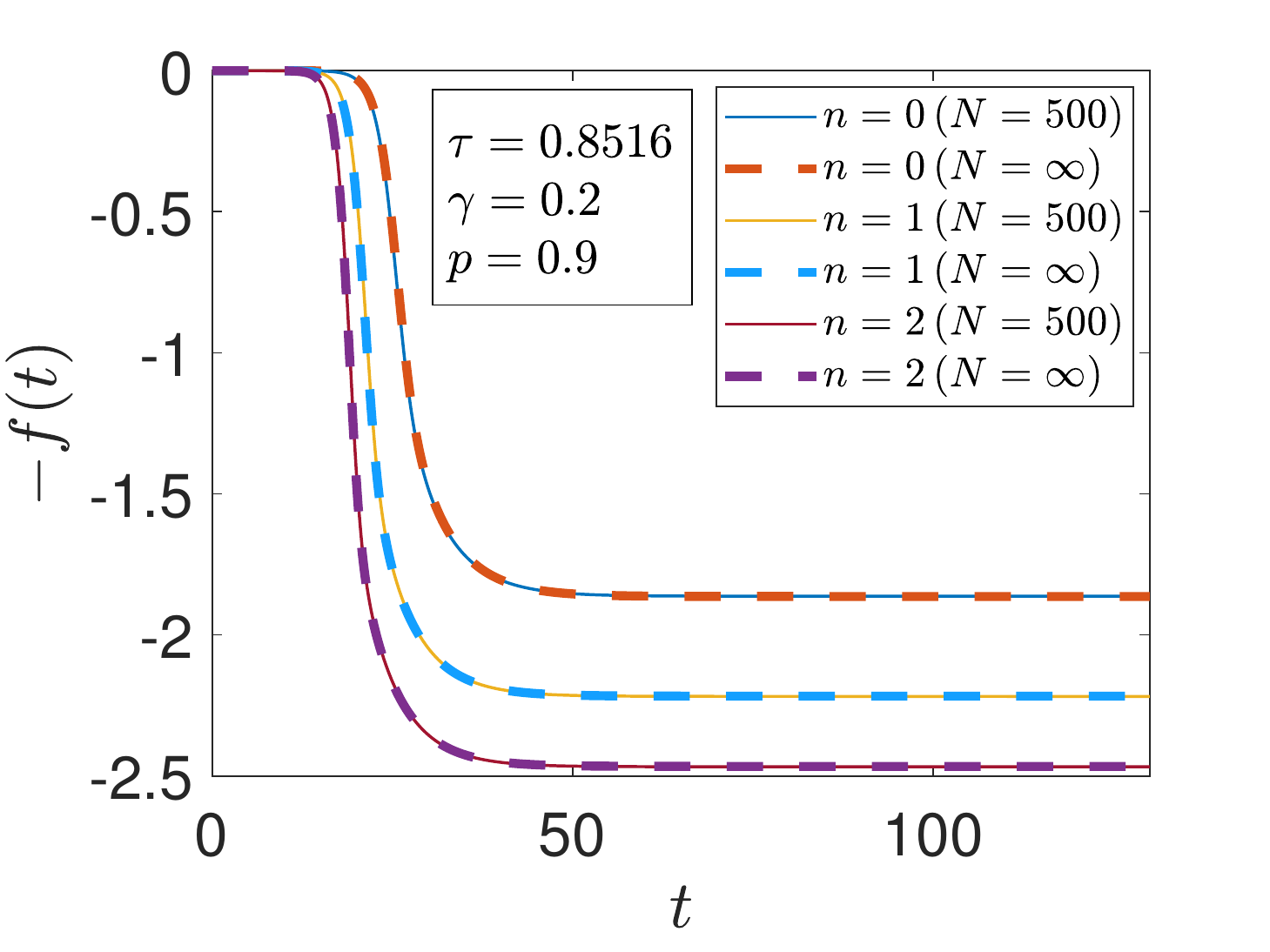}\label{case3b}}
	\end{center}
	\caption{Comparison between the continuum solution $-f(t)$ of Eq.~(\ref{h4}) and the discrete network solution
		(Eqs~(\ref{Sdot2}) and (\ref{Idot2})) with $N=500$ fitted using  \textbf{$\hat{S}(t)=\hat{S}(-\nu) \circ e^{-f(t)\sqrt{\hat{\beta}}}$}. (a) $\hat{U}(\hat{\beta})=\frac{1}{b^{a}\Gamma(a)}\hat{\beta}^{(a-1)}e^{\frac{-\hat{\beta}}{b}}$  and $\hat{V}(\hat{\beta})=\hat{U}(\hat{\beta})$. (b)  $\hat{U}(\hat{\beta})=r_n \cdot ( 1+\hat{\beta}^n),\,0\le \hat{\beta} \le 6$  and  $\hat{V}(\hat{\beta})$ randomized as in Fig
		\ref{Figure3}. The history function for Eq.~(\ref{h4}) is taken as $f(t)=\omega\times(1+\frac{t}{\nu})$, with $\omega$ is adjusted to match the results from (Eqs~(\ref{Sdot2}) and (\ref{Idot2})). Other numerical parameters are mentioned in text boxes within subplots.}
	\label{Figure4}
\end{figure*}

Motivated by the foregoing observations from numerical simulations, we now look for solutions of the form
$$S(\beta,t) = \phi(\beta) e^{-f(t) \sqrt{\beta}},$$
where $\phi(\beta)$ can be determined from initial conditions.
Substituting into Eq.~(\ref{e3}) yields
\begin{eqnarray} 
\label{f1}
&-& \sqrt{\beta} \phi(\beta) \dot f(t) e^{-f(t) \sqrt{\beta}} =\nonumber\\
&- &\sqrt{\beta} \phi(\beta)  e^{-f(t) \sqrt{\beta}} \int_0^{\infty} \sqrt{\xi} I(\xi,t)\, d \xi.
\end{eqnarray}
The integral on the right hand side is purely a function of time $t$: let us call it $g(t)$. Then we have
\begin{equation} \label{f2}
\dot f(t)  = g(t).
\end{equation}
Also, Eq.~(\ref{e4}) becomes
\begin{eqnarray} 
\label{f3}
\dot I(\beta,t) &=& \sqrt{\beta} \phi(\beta) e^{-f(t-1) \sqrt{\beta}} g(t-1) - \gamma I(\beta,t) \nonumber\\
&-& \bar p \sqrt{\beta}\phi(\beta) e^{-f(t-\nu) \sqrt{\beta}} g(t-\nu).
\end{eqnarray}
We multiply both sides above by $\sqrt{\beta}$ and integrate with respect to $\beta$.
Defining
\begin{equation} \label{f4}
H(t) = \int_0^{\infty} \beta \phi(\beta) e^{-f(t) \sqrt{\beta}} d \beta,
\end{equation}
and assuming the integral and the derivative on the left hand side can be interchanged, Eq~(\ref{f3}) becomes
\begin{eqnarray}
\label{f5}
\dot g(t) &=& H(t-1) g(t-1)  - \gamma g(t)\nonumber\\
&-& \bar p H(t-\nu) g(t-\nu).
\end{eqnarray}
Any solution of Eqs.~(\ref{f2}) and (\ref{f5}), with $H(t)$ defined by Eq.~(\ref{f4}), corresponds to an {\em exact} solution of Eqs~(\ref{e3}) and (\ref{e4}). Differentiating Eq.~(\ref{f2}) with respect to time and substituting for $\dot{g}(t)$ from Eq.~(\ref{f5}), we obtain:
\begin{eqnarray}
\label{h1a}
\ddot f(t) &=& H(t-1) \dot f(t-1)- \gamma \dot f(t)\nonumber\\
&-& \bar p H(t-\nu) \dot f(t-\nu).
\end{eqnarray}
Separately, multiplying both sides Eq.~(\ref{f4}) by $\dot f(t)$, we obtain
\begin{equation}
H(t) \dot f(t) = \int_0^{\infty} \beta \phi(\beta) e^{-f(t) \sqrt{\beta}} \dot f(t) d \beta,
\end{equation}
where we notice that the right hand side is integrable with respect to time, giving
\begin{equation} \label{h3}
\int H(t) \dot f(t) dt = - \int_0^{\infty} \sqrt{\beta} \phi(\beta)  e^{-f(t) \sqrt{\beta}} d \beta + C,
\end{equation}
where $C$ is a time-independent constant (it does not depend on $\beta$ either, because $\beta$ here is a dummy variable of integration).
Let the integral on the right hand side of Eq~(\ref{h3}) be called $G(f(t))$, i.e.,
\begin{equation} \label{h3a}
G(f(t)) =\int_0^{\infty} \sqrt{\beta} \phi(\beta)  e^{-f(t) \sqrt{\beta}} d \beta,
\end{equation}
where we notice that the time $t$ within this definition is merely a parameter that determines $f$, and so we can also simply write
\begin{equation} \label{h3aa}
G(f) =\int_0^{\infty} \sqrt{\beta} \phi(\beta)  e^{-f \sqrt{\beta}} d \beta
\end{equation}
if it suits our purpose.
Equation~(\ref{h1a}) becomes, upon one integration,
\begin{equation} \label{h4}
\dot f(t) = -G(f(t-1)) + \bar p G(f(t-\nu))  - \gamma f(t) + C_0,
\end{equation}
where $C_0$ is a constant of integration and can be evaluated by setting $f(t)=0$ when $\dot{f}(t)=0$ (at the time of infinitesimal intiation of infection), and is given by
\begin{equation} 
\label{C0}
C_0 =(1-\bar{p})\int_0^{\infty} \sqrt{\beta} \phi(\beta) d \beta.
\end{equation}

For many initial distribution functions $\phi(\beta)$, $G(f(t))$ and $C_0$ can be evaluated in closed form. For example, for the uniform distribution
$$\phi(\beta)=\frac{1}{B}, \quad  0\le \beta \le B,$$
we have
$$G(f(t))=K(f(t))R(f(t))$$ where
$$K(f(t))=-\frac{2{\rm e}^{-\sqrt{B}f(t)}}{B^{1/2}f(t)^{3}},$$ 
$$R(f(t))=-2{\rm e}^{\sqrt{B}f(t)}B^{3/2}+f(t)^{2}B^{5/2}+2f(t)B^{2}+2B^{3/2},$$
and $$C_0=(1-\bar{p})\frac{2\sqrt{B}}{3}.$$

In this way, at least in principle (even if the integrals cannot be evaluated in closed form), Eq.~(\ref{h4}) is a {\em first order scalar DDE} that governs the dynamics of the continuum limit of our network. The infinite-dimensional variability with respect to the distributed parameter $\beta$ has collapsed into a single dimension. This observation of complete dimensional collapse is one of the main contributions of the paper. 
Our result is independent of the distribution $\phi(\beta)$.

Figure~\ref{Figure4} compares the continuum solution $-f(t)$ obtained from Eq.~(\ref{h4}), and the finite network solution
(Eqs~(\ref{Sdot2}) and (\ref{Idot2}))
fitted to \textbf{$\hat{S}(t)=\hat{S}(-\nu)\circ e^{-f(t)\sqrt{\hat{\beta}}}$} for different parameter values (see figure caption for details). The results match closely.

	From Figure~\ref{Figure4}, it is evident that, if the initial conditions for the infected population are all sufficiently small, the subsequent evolution does not depend very much on them. This is because the interaction term smoothes out all those initial variations, and the dimensional collapse occurs, which is the main point of the paper. In particular see the simulation results in Fig.~\ref{Figure4}(a) and Fig.~\ref{Figure4}(b). In one case the infective population is taken to have a smooth variation with respect to $\beta$, and in the other case it is a random variation with respect to $\beta$. The results are essentially the same, and the match is nearly perfect. 

We now proceed to interpret $G(f)$ in terms
of the moment generating function of an underlying randomly distributed quantity,
\begin{equation}
\sqrt{\beta} = u,
\end{equation}
by defining an intermediate quantity (using an overbar to formally distinguish the two functions)
\begin{equation}
\phi(\beta) = \bar \phi(u).
\end{equation}
For example, if $\phi(\beta) = \beta$, then we mean $\bar \phi(u) = u^2$.

Equation~(\ref{h3a}), written as Eq.\ (\ref{h3aa}), becomes
\begin{equation} \label{h5}
G(f) = 2 \int_0^{\infty} u^2 \bar \phi(u)  e^{-f u} d u.
\end{equation}
Now, if we interpret $\phi(\beta)$ as the probability density function of the random variable $\beta$ in the population, and think of $\psi(u)$ as the probability density function of the transformed variable $u =\sqrt{\beta}$ in the same population, then it must be true that
\begin{equation}
\label{transf}
\phi(\beta) = \bar \phi(u) = \frac{\psi(u)}{2u},
\end{equation}
which gives
\begin{equation} \label{h6ab}
G(f) = \int_0^{\infty} u  \psi(u)  e^{-f u} d u,
\end{equation}
where we see that, except for a sign change in $f$, $G(f)$ is the first derivative with respect to $f$ of the moment generating function~\cite{hogg2005introduction} of the distributed quantity $u = \sqrt{\beta}$.

If we expand in a series for small $f$, we obtain
\begin{equation} \label{h6}
G(f) = m_1 -  m_2 f + m_3 \frac{f^2}{2!}- m_4 \frac{f^3}{3!}+ \cdots,
\end{equation}
where the coefficients $m_k$ are moments of $u$, given by
\begin{equation}\label{moments}
m_{k}=\intop_{0}^{\infty}u^{k}\psi(u)du,
\end{equation}
assuming these moments exist.
In other words, $m_1$ is the population mean of $\sqrt{\beta}$, $m_2$ is the population mean of $\beta$, $m_3$ is the population mean of $\beta^{3/2}$, and so on. Returning to Eq.~(\ref{h4}) and expanding for small $f$, and retaining up to cubic terms (i.e., retaining up to second order nonlinear correction terms), we have
\begin{eqnarray} 
\label{h7}
\dot f(t) &=& \left(-m_1+  m_2 f(t-1) - m_3 \frac{f(t-1)^2}{2!}+ m_4 \frac{f(t-1)^3}{3!}\right)\nonumber\\
&+ & \bar p \left ( m_1- m_2 f(t-\nu) + m_3 \frac{f(t-\nu)^2}{2!}- m_4 \frac{f(t-\nu)^3}{3!} \right )\nonumber\\
&-& \gamma f(t) +C_0. 
\end{eqnarray}
In Eq.~(\ref{h7}), if we are interested in an outbreak that starts from an infinitesimal infection, such that $f=0$ when $\dot f = 0$, then the constant
$$C_0=m_1(1-\bar p),$$
consistent with Eq.\ (\ref{C0}), leaving  
\begin{eqnarray}
\label{h8}
\dot f(t) &=& \left( m_2 f(t-1) - m_3 \frac{f(t-1)^2}{2!}+ m_4 \frac{f(t-1)^3}{3!}\right)\nonumber\\
&+ &\bar p \left (- m_2 f(t-\nu) + m_3 \frac{f(t-\nu)^2}{2!}- m_4 \frac{f(t-\nu)^3}{3!} \right )\nonumber\\
&-& \gamma f(t).
\end{eqnarray}

Linearising Eq.~(\ref{h8}) for small $f$, we obtain
\begin{equation}
\label{h9}
\dot f(t)= m_2 f(t-1) -\bar p  m_2 f(t-\nu)- \gamma f(t).
\end{equation}
Substituting $f(t)=e^{st}$, we find
\begin{equation}
s=m_2e^{-s}-m_2\bar{p}e^{-\nu s}-\gamma
\end{equation}
This characteristic equation matches one studied in \cite{young2019consequences,Vyas2020}, and the stability condition is known to be
(recall Eq.\ (\ref{pbareq}))
\begin{equation}
\label{stabresult}
R_0=m_2\frac{(1-\bar{p})}{\gamma}<1.
\end{equation}
In fact, since $m_2$ is the second moment of $\sqrt{\beta}$, i.e., the population average of $\beta$, this validates the use of an average
value in the lumped model (Eqs.~(\ref{SdotR})-(\ref{RdotR})), studied in \cite{young2019consequences,Vyas2020}. Since our continuum model
allows a general distribution for $\beta$, we can  consider the specific case of a Dirac delta function (i.e., $\beta$ is not random)
\begin{equation}
\label{av}
\psi(u)=\delta\left(u-\sqrt{\beta_{0}}\right).
\end{equation}
Equation (\ref{h6ab}) yields
\begin{equation} \label{h6a}
G(f) = \sqrt{\beta_0}\,  e^{-f \sqrt{\beta_0}}.
\end{equation}
Equation (\ref{h4}) becomes (inserting $C_0$ to match initial conditions of $f=0$ when $\dot f = 0$)
$$
\dot f(t) = -\sqrt{\beta_0}\,  e^{-f(t-1) \sqrt{\beta_0}} + \bar p \sqrt{\beta_0}\,  e^{-f(t-\nu) \sqrt{\beta_0}}  \cdots
$$
$$\quad \quad - \gamma f(t) + \sqrt{\beta_0}(1 -\bar p).
$$
Setting $f(t) = \sqrt{\beta_0}P(t)$, we obtain
\begin{equation} \label{eq15old}
\dot P(t) =  \bar p   e^{-\beta_0 P(t-\nu) } -  e^{-\beta_0 P(t-1)} - \gamma P(t) + 1 - \bar p,
\end{equation}
which exactly matches Eq.~(15) of~\cite{Vyas2020}.

Further, even for variable $\beta$, the correpondence between Eq.\ (\ref{h4}) and Eq.\ (\ref{eq15old}) actually holds up to second order.
If the expansion for small $f$ in Eq.\ (\ref{h8}) is truncated at second order, i.e., we retain only the second and third moments of $u$, then it is easy to show that
a simple scaling of $f$ makes that equation match the second order expansion of Eq.\ (\ref{eq15old}). The implication is that, for relatively small outbreaks, the dynamics under distributed $\beta$ (i) depends on the expected values of $\beta$ and $\beta^{3/2}$, and (ii) is the same, up to a linear
scaling, as the dynamics with a single fixed $\beta$ as studied in \cite{Vyas2020}. For larger outbreaks, higher moments of the distribution of $\beta$
begin to play a role, and 
the match with Eq.\ (\ref{eq15old}) deteriorates.

\begin{figure*}[htpb!]
	\begin{center}
		\subfigure[]{\includegraphics[width=0.45\textwidth]{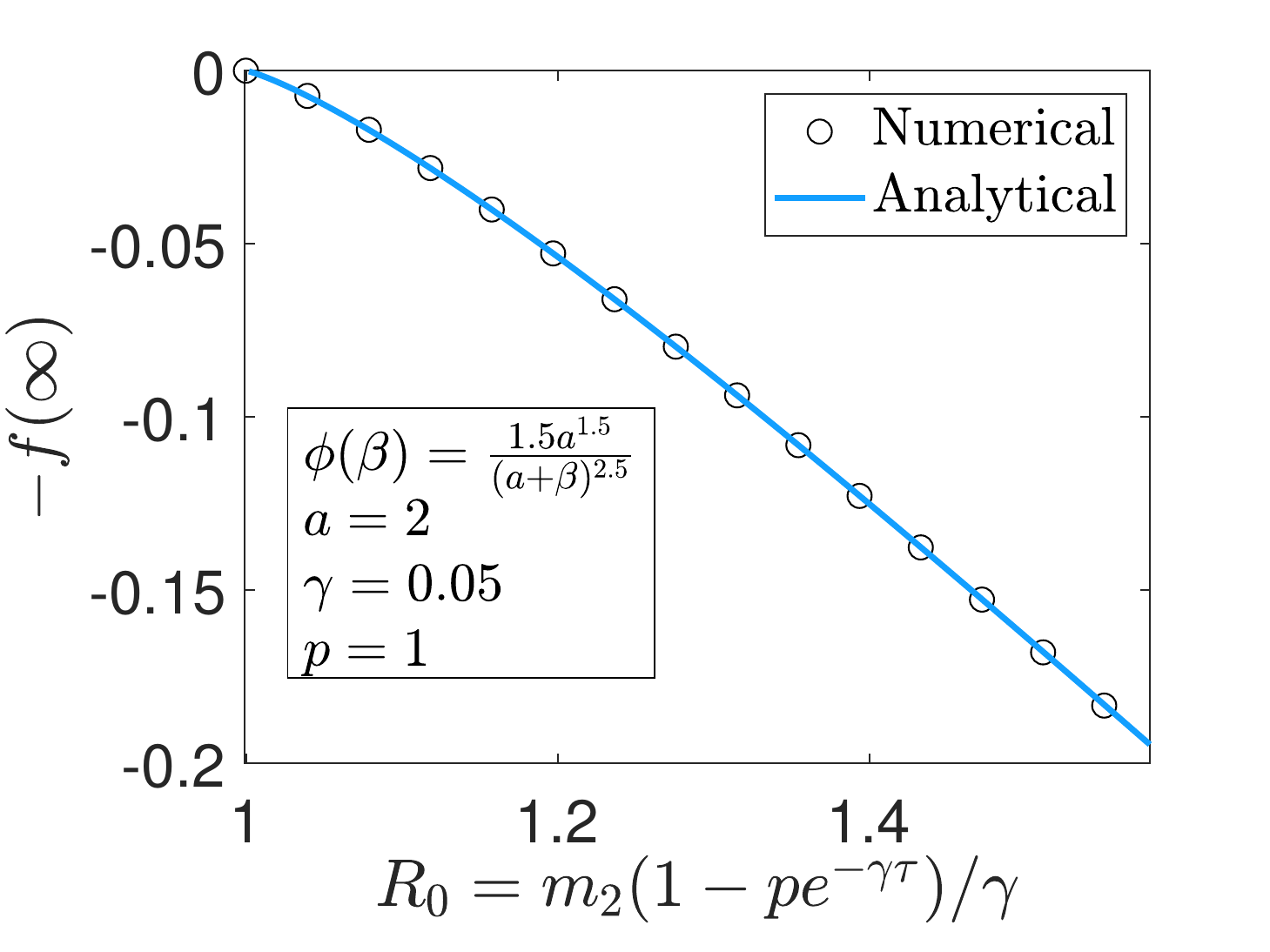}\label{case7}}
		p		\subfigure[]{\includegraphics[width=0.45\textwidth]{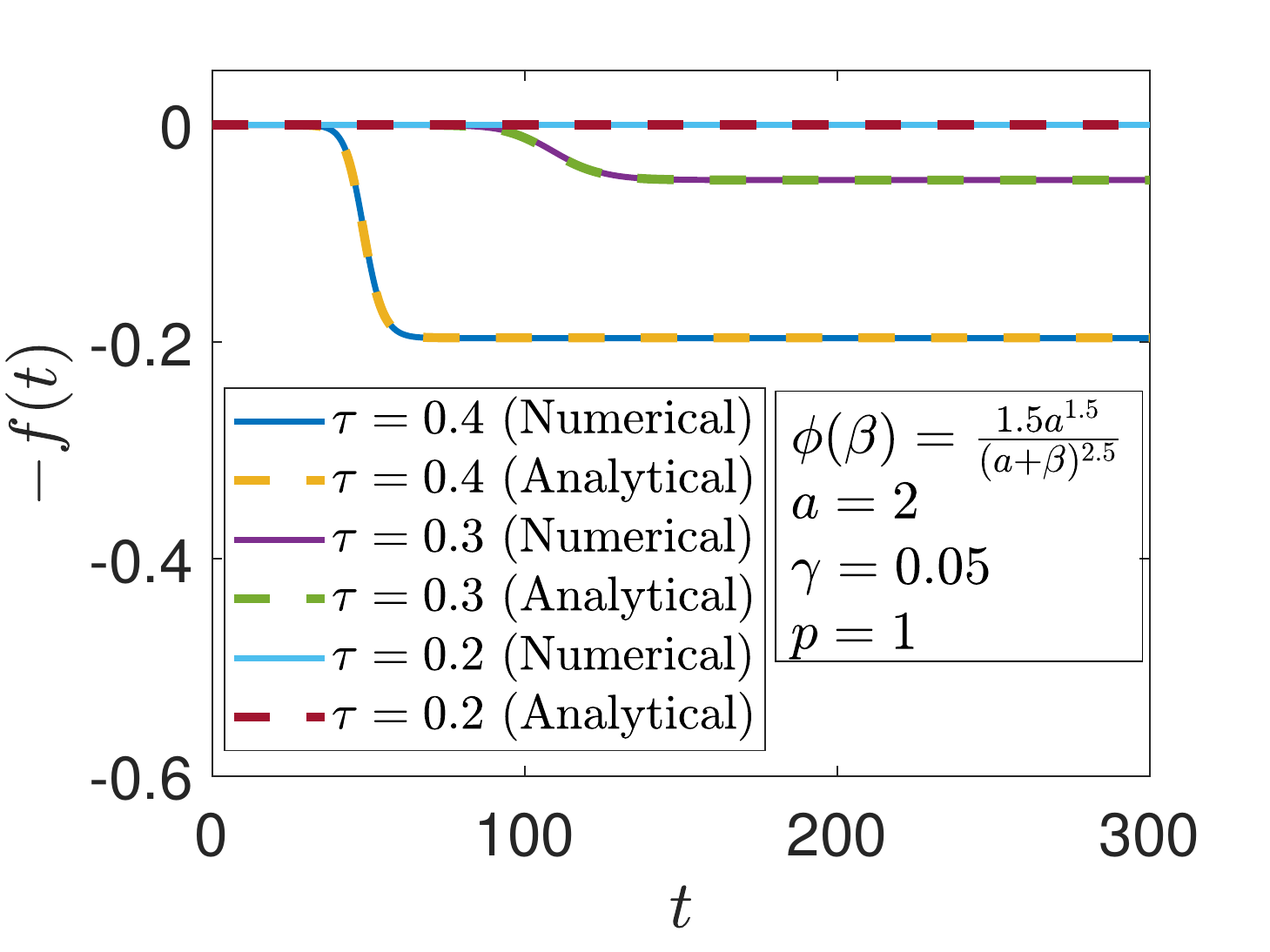}\label{case8}}
	\end{center}
	\caption{ An example with unbounded moment $m_3$ (see Eq.\ (\ref{infmom})). (a) $R_0$ on the horizontal axis governs stability.
		Each circle, labeled ``numerical'', represents the saturation value observed in a numerical solution based on integration of Eq.~(\ref{h4}) combined with Eq.~(\ref{h6ab}). At every time step, the integral in Eq.~(\ref{h6ab}) is evaluated numerically
		for the truncated domain $0 < u < 1000$. The continuous line labeled ``analytical'' is from Eq.~(\ref{h4}) with Eq.~(\ref{TE}) used directly. (b) Comparison of time responses $-f(t)$, with ``numerical'' and ``analytical'' implying the same as in (a). The initial function used was $f(t)=1\times10^{-9}+1\times10^{-9}\left(1+\frac{t}{\nu}\right),\,t\le 0$.}
	\label{figure6}
\end{figure*}

Finally, we close with an example where the required moments in Eq.\ (\ref{h8}) do not all exist. Let the probability density function of $\beta$ be
\begin{equation}
\label{infmom}
\phi(\beta) = \frac{3 \cdot a^{3/2}}{2(a+\beta)^{5/2}}, \quad a>0,
\end{equation}
for which,  in Eq.\ (\ref{h8}), $m_2 = 2a$ and $m_3 = \infty$. This distribution has a long tail.
The integral $G(f)$ for $f>0$ (recall Eq.\ (\ref{h6})) can be found (using Maple) in terms of special functions, and the first few terms of an asymptotic series for small $f>0$ is as shown below:
\begin{eqnarray}
\label{TE}
G(f) &=& \sqrt{a} - 2af \nonumber\\
&-& \frac{3a^{3/2}f^2}{2} \left (\gamma_E -\frac{1}{6} - \ln 2 + \frac{\ln a}{2} + \ln f   \right ) \nonumber\\
& +& \frac{4a^2 f^3}{3}\nonumber\\
&+& \frac{5a^{5/2}f^4}{16} \left (\gamma_E
-\frac{21}{20} - \ln 2 + \frac{\ln a}{2} +  \ln f  \right )\nonumber+ \cdots\\
\end{eqnarray}
where $\gamma_E = 0.57721566\cdots$ is Euler's constant, also called the Euler-Mascheroni constant. The series has an $f^2 \ln f$ term, reflecting the unbounded third moment of $u$. Equation~(\ref{TE}) can be used in a numerical solution of the DDE in Eq.\ (\ref{h4}) with Eq.\ (\ref{C0}), provided the outbreak is not too big. For comparison, straightforward solution of the governing equations can be tried with direct numerical integration of Eq.~(\ref{h4}) combined with Eq.~(\ref{h6ab}) truncated at some suitably large value of $u$ (we used $u=1000$). Some such solutions are shown in Fig \ref{figure6}. The match is essentially perfect, which indicates that finiteness of moments of $u$, and hence the moments of $\beta$, is not needed for the underlying DDE to have useful solutions.

\section{Discussion and Conclusions}
In this paper we have considered the extension of a compartmental model for an epidemic or pandemic to a network model, structured on the infectivity parameter $\beta$ rather than background factors like age or occupation. We have considered a simple interaction model between susceptible individuals with different infectivity parameters $\beta_k$ and $\beta_r$, and examined the case where the effective interaction parameter
between these groups is monotonic in each parameter, symmetric, and separable. Under reasonable qualitative requirements, we settled on
an interaction term
of the form $\sqrt{\beta_k \beta_r}$. With this interaction, the continuum limit of the network model yielded a pair of delayed integrodifferential equations parameterized by $\beta$. However, numerical solutions also suggested a remarkable dimensional collapse in the underlying dynamics,
such that the entire variation due to the continuously distributed $\beta$ could be captured using a single function of time. The dynamics in terms of this function, $f(t)$, turned out to be given by a single first order nonlinear DDE. We have referred to this
dramatic simplification as {\em complete dimensional collapse}.

We are aware of similar dimensional collapse for the Fokker-Planck equation for a plasma in an ion trap (a parametrically forced system without delays): see \cite{bhattacharjee2019unifying} and references
therein. However, we are not aware of such collapse being previously noted for any models within mathematical epidemiology.

The DDE governing the new variable $f(t)$ has strong connections with the lumped-model DDE where $\beta$ is assumed to be a single constant for the entire population. When the first few moments of the underlying parameter $u = \sqrt{\beta}$ in the population are finite, a direct correspondence up to second order can be established with the single-$\beta$ equation. Even if the required moments are not finite, the DDE remains well posed, although its small-$f$ behavior may be more complicated.

There are interesting social implications that emerge from our model. For example, in the finite-moments case, even if $\beta$ is distributed such that a small proportion of the population has rather high $\beta$ values, there is no immediate catastrophic growth because that subpopulation, being small, tends to interact mostly with other parts of the population whose $\beta$ values are smaller. Even in a population where some people engage in behavior corresponding to high $\beta$ (i.e., not obeying social distancing), low-$\beta$ behavior from {\em other} members of the population can help to both limit the pandemic and provide differential benefits based on adopted risk levels. Satisfyingly, the criterion for unstable growth from small initial infection is governed solely by the population average of $\beta$, although the final saturation value of the infection depends on higher moments. In particular, for a small outbreak due to weak instability under infinitesimal initiation, the saturation value is essentially determined by the population averages of $\beta$ and $\beta^{3/2}$, or of $u^2$ and $u^3$ respectively.

Future work with such continuum models may incorporate additional features for greater realism, including non-separable interaction terms, behavior modification by the population as the outbreak progresses, new approximate solutions for various special cases including cases with unbounded moments of $u$, and direct comparisons of model predictions with national or international data.

\section*{Acknowlegements}
We thank Sankalp Tiwari for commenting on this work.

\section*{Funding}
We have not received any financial support for conducting this research.

\section*{Compliance with ethical standards}

\textbf{Conflict of Interest:} The authors declare that they have no conflict of interest.


\begin{thebibliography}{30}
	
	
	\bibitem{kermack1927contribution}
	Kermack, W.O., McKendrick, A.G.: A contribution to the mathematical theory of
	epidemics.
	\newblock Proceedings of the Royal Society of London A: Mathematical, Physical
	and Engineering Sciences \textbf{115}(772), 700--721 (1927)
	
	\bibitem{hethcote2009basic}
	Hethcote, H.W.: The basic epidemiology models: models, expressions for {R0},
	parameter estimation, and applications.
	\newblock In: Mathematical Understanding of Infectious Disease Dynamics, pp.
	1--61. World Scientific (2009)
	
	
	\bibitem{brauer2012mathematical}
	Brauer, F., Castillo-Chavez, C., Castillo-Chavez, C.: Mathematical models in
	population biology and epidemiology, Texts in Applied Mathematics, vol.~2.
	\newblock Springer-Verlag, New York (2012)
	
	\bibitem{hethcote2000mathematics}
	Hethcote, H.W.: The mathematics of infectious diseases.
	\newblock SIAM Review \textbf{42}(4), 599--653 (2000)
	
	
	\bibitem{gerberry2009seiqr}
	Gerberry, D.J., Milner, F.A.: An {SEIQR} model for childhood diseases.
	\newblock Journal of Mathematical Biology \textbf{59}(4), 535--561 (2009)
	
	\bibitem{hethcote1989three}
	Hethcote, H.W.: Three basic epidemiological models.
	\newblock In: Applied Mathematical Ecology, pp. 119--144. Springer (1989)
	
	
	\bibitem{zuzek2015epidemic}
	Zuzek, L.A., Stanley, H.E., Braunstein, L.A.: Epidemic model with isolation in
	multilayer networks.
	\newblock Scientific Reports \textbf{5}, 12151 (2015)
	
	
	\bibitem{morita2016six}
	Morita, S.: Six susceptible-infected-susceptible models on scale-free networks.
	\newblock Scientific Reports \textbf{6}, 22506 (2016)
	
	
	\bibitem{hasegawa2017efficiency}
	Hasegawa, T., Nemoto, K.: Efficiency of prompt quarantine measures on a
	susceptible-infected-removed model in networks.
	\newblock Physical Review E \textbf{96}(2), 022311 (2017)
	
	
	\bibitem{strona2018rapid}
	Strona, G., Castellano, C.: Rapid decay in the relative efficiency of
	quarantine to halt epidemics in networks.
	\newblock Physical Review E \textbf{97}(2), 022308 (2018)
	
	
	\bibitem{coelho2008epigrass}
	Coelho, F.C., Cruz, O.G., Code{\c{c}}o, C.T.: Epigrass: a tool to study disease
	spread in complex networks.
	\newblock Source Code for Biology and Medicine \textbf{3}(1), 3 (2008)
	
	\bibitem{keeling2005networks}
	Keeling, M.J., Eames, K.T.: Networks and epidemic models.
	\newblock Journal of the Royal Society Interface \textbf{2}(4), 295--307 (2005)
	
	
	\bibitem{singh2020age}
	Singh, R., Adhikari, R.: Age-structured impact of social distancing on the
	{COVID}-19 epidemic in {I}ndia.
	\newblock arXiv preprint, arXiv:2003.12055  (2020)
	
	\bibitem{barbera2013spread}
	Barbera, E., Consolo, G., Valenti, G.: Spread of infectious diseases in a
	hyperbolic reaction-diffusion susceptible-infected-removed model.
	\newblock Physical Review E \textbf{88}(5), 052719 (2013)
	
	\bibitem{ruan2007spatial}
	Ruan, S.: Spatial-temporal dynamics in nonlocal epidemiological models.
	\newblock In: Mathematics for Life Science and Medicine, pp. 97--122.
	Springer-Verlag Berlin Heidelberg (2007)
	
	\bibitem{jones2009differential}
	Jones, D.S., Plank, M., Sleeman, B.D.: Differential equations and mathematical
	biology.
	\newblock CRC press (2009)
	
	\bibitem{schneckenreither2008modelling}
	Schneckenreither, G., Popper, N., Zauner, G., Breitenecker, F.: Modelling
	{SIR}-type epidemics by {ODE}s, {PDE}s, difference equations and cellular
	automata--{A} comparative study.
	\newblock Simulation Modelling Practice and Theory \textbf{16}(8), 1014--1023
	(2008)
	
	
	\bibitem{medlock2003spreading}
	Medlock, J., Kot, M.: Spreading disease: integro-differential equations old and
	new.
	\newblock Mathematical Biosciences \textbf{184}(2), 201--222 (2003)
	
	
	\bibitem{van2002time}
	Van~den Driessche, P.: Time delay in epidemic models.
	\newblock IMA Volumes In Mathematics and its Applications \textbf{125},
	119-128 (2002)
	
	
	\bibitem{young2019consequences}
	Young, L.S., Ruschel, S., Yanchuk, S., Pereira, T.: Consequences of delays and
	imperfect implementation of isolation in epidemic control.
	\newblock Scientific Reports \textbf{9}(1), 1--9 (2019)
	
	
	\bibitem{bocharov2000numerical}
	Bocharov, G.A., Rihan, F.A.: Numerical modelling in biosciences using delay
	differential equations.
	\newblock Journal of Computational and Applied Mathematics \textbf{125}(1-2),
	183--199 (2000)
	
	\bibitem{nelson2002mathematical}
	Nelson, P.W., Perelson, A.S.: Mathematical analysis of delay differential
	equation models of {HIV}-1 infection.
	\newblock Mathematical Biosciences \textbf{179}(1), 73--94 (2002)
	
	
	\bibitem{alexander2008delay}
	Alexander, M.E., Moghadas, S.M., R{\"o}st, G., Wu, J.: A delay differential
	model for pandemic influenza with antiviral treatment.
	\newblock Bulletin of Mathematical Biology \textbf{70}(2), 382--397 (2008)
	
	\bibitem{gourley2008dynamics}
	Gourley, S.A., Kuang, Y., Nagy, J.D.: Dynamics of a delay differential equation
	model of {H}epatitis {B} virus infection.
	\newblock Journal of Biological Dynamics \textbf{2}(2), 140--153 (2008)
	
	\bibitem{rakkiyappan2019}
	Rakkiyappan, R., Latha, V.P., Rihan, F.A.: A fractional-order model for {Z}ika
	virus infection with multiple delays.
	\newblock Complexity p. 4178073
	
	
	\bibitem{Vyas2020}
	Vyasarayani, C.P., Chattejee, A.: New approximations, and policy implications,
	from a delayed dynamic model of a fast pandemic.
	\newblock arXiv preprint, arXiv:2004.03878  (2020)
	
	\bibitem{Vyas20202}
	Vyasarayani, C.P., Chattejee, A.: Complete dimensional collapse in the
	continuum limit of a delayed {SEIQR} network model with separable distributed
	infectivity.
	\newblock arXiv preprint, arXiv:2004.12405  (2020)
	
	
	\bibitem{blackwood2018introduction}
	Blackwood, J.C., Childs, L.M.: An introduction to compartmental modeling for
	the budding infectious disease modeler.
	\newblock Letters in Biomathematics \textbf{5}(1), 195--221 (2018)
	
	
	\bibitem{hogg2005introduction}
	Hogg, R.V., McKean, J., Craig, A.T.: Introduction to mathematical statistics.
	\newblock Pearson Education (2005)
	
	\bibitem{bhattacharjee2019unifying}
	Bhattacharjee, A., Shah, K., Chatterjee, A.: Unifying averaged dynamics of the
	fokker-planck equation for paul traps.
	\newblock Physics of Plasmas \textbf{26}(1), 012302 (2019)
	
\end{thebibliography}
\end{document}